\documentclass[11pt]{article}
\usepackage[utf8]{inputenc}
\usepackage{amsmath,amssymb,amsfonts,amstext}
\usepackage{cite}
\usepackage{fancyhdr}
\usepackage[font=footnotesize,labelfont=bf,justification=centerlast,width=.94\textwidth]{caption}
\usepackage{mathrsfs}
\usepackage{bbold}
\usepackage[dvipsnames]{xcolor}

\usepackage[colorlinks=true,pdfstartview=FitV,linkcolor= BrickRed,citecolor= BrickRed,urlcolor= BrickRed,hyperindex=true,hyperfigures=true]{hyperref}
\hypersetup{linktoc=page}

\makeatletter \@addtoreset{equation}{section} \makeatother

\makeatletter
\g@addto@macro\bfseries{\boldmath}
\makeatother

\begin{document}
\begin{titlepage}
\begin{flushright} 
\end{flushright}
\vfill
\begin{center}
{\Large \textbf{A Worldsheet Approach to $\mathbf{\mathcal{N}=1}$ Heterotic Flux Backgrounds}}
\vskip 1cm 
Dan Isra\"el and Yann Proto\vskip 1cm 
\textit{Sorbonne Universit\'{e}, CNRS, Laboratoire de Physique Th\'{e}orique et Hautes \'{E}nergies, LPTHE, F-75005 Paris, France}
\vskip 0.5cm 
\textit{E-mail:} \texttt{israel@lpthe.jussieu.fr}, \texttt{yproto@lpthe.jussieu.fr}
\end{center}
\vfill
\begin{center} \textbf{Abstract} \end{center}
\begin{quote}
Heterotic backgrounds with torsion preserving minimal supersymmetry in four dimensions can be obtained as orbifolds of principal $T^{2}$ bundles over $K3$. We  consider a worldsheet description of these backgrounds as gauged linear sigma-models (GLSMs) with $(0,2)$ supersymmetry. Such a formulation provides a useful framework in order to address the resolution of singularities of the orbifold geometries. We investigate the constraints imposed by discrete symmetries on the corresponding torsional GLSMs. In particular, the principal $T^{2}$ connection over $K3$ is inherited from $(0,2)$ vector multiplets. As these vectors gauge global scaling symmetries of products of projective spaces, the corresponding $K3$ geometry is naturally realized as an algebraic hypersurface in such a product (or as a branched cover of it). We outline the general construction for describing such orbifolds. We give explicit constructions for automorphisms of order two and three.
\end{quote}
\vfill
\end{titlepage}

\tableofcontents

\section{Introduction}
Heterotic flux compactifications provide a very natural framework in order to build phenomenological models within string theory. With such compactifications, grand unified gauge groups can be incorporated easily, while the question of moduli stabilization is partly addressed by turning on fluxes along the internal manifold. Heterotic flux backgrounds can also lead us to a better understanding of string vacua, as the quantum geometry of string compactifications outside the Calabi--Yau landscape remains largely unknown.

Familiar Calabi--Yau compactifications come with several facilitating characteristics. Theorems of Yau~\cite{Yau:1978cfy} and Donaldson--Uhlenbeck--Yau~\cite{Donaldson:1985zz,Uhlenbeck:1986sl} ensure the existence of classical solutions to the heterotic equations of motion, when there is no topological obstruction. Moreover, such backgrounds admit a large volume limit in the moduli space, which allows for a perturbative treatment parametrized by the radius of the internal manifold. With this perturbative expansion, $\alpha'$ corrections to the field equations can be processed order by order, in order to extend classical solutions to the exact string solutions~\cite{Witten:1985bz,Witten:1986kg}.

These features are lost for the most part when turning to heterotic flux compactifications, which are associated with non-K\"{a}hler geometries. Beyond the K\"{a}hler realm, some crucial aspects disappear, such as Hodge decomposition or special geometry. The aforementioned theorems do not extend to torsional spaces, and existence of solutions has to be proven case by case. A major technical challenge also comes from the Bianchi identity. For non-K\"{a}hler manifolds, it becomes highly non-linear in the flux. Moreover, as the Bianchi identity does not scale homogeneously under rescalings of the internal metric, some of the cycles of the internal manifold are typically fixed at $\alpha'$ size. The corresponding two-dimensional worldsheet theory is thus strong coupled, and the validity of the effective supergravity can be questioned.
A worldsheet approach becomes inevitable to probe the theory.

While conditions for a four-dimensional heterotic supergravity compactification to preserve minimal $\mathcal{N}=1$ supersymmetry are well understood~\cite{Hull:1986kz,Strominger:1986uh}, a very limited number of such backgrounds is known to exist. The main class of solutions consists of a principal $T^{2}$ bundle over a $K3$ surface. The particular structure of these spaces allows to circumvent most of the difficulties that come with non-K\"{a}hler compactifications. Dual M-theory constructions~\cite{Dasgupta:1999ss} lead to believe that, even if the $\alpha'$ expansion of such solutions is to be taken with caution, their existence should still be guaranteed. Furthermore, while the torus cycles of these solutions are quantized in $\alpha'$, the base volume can be taken arbitrarily large. Consequently, such compactifications admit a good eight-dimensional large volume limit, which corresponds to the heterotic theory compactified on $T^{2}$.

For this class of torsional backgrounds one can identify eight supercharges in space-time.\footnote{Turning on some components of the curvature of the torus bundle reduces by half the number of supersymmetry preserved by the geometry\cite{Becker:2007ea}; they have no known worldsheet realization in the formalism that we use in this work.} Extended spacetime supersymmetry imposes strong constraints on the worldsheet superconformal field theory. Maintaining $\mathcal{N}=2$ spacetime supersymmetry requires that the internal right-moving superconformal algebra, with central charge $\Bar{c}=9$, splits into a $(0,4)$ piece with $\Bar{c}=6$ and a free $(0,2)$ piece with $\Bar{c}=3$ and two $U(1)$ currents \cite{Banks:1988yz}. If such a theory is to be constructed geometrically,\footnote{The analysis of~\cite{Melnikov:2010pq} assumes a smooth geometry for the target space.} then the corresponding target space must be a $T^{2}$ bundle over a $K3$ base \cite{Melnikov:2010pq}. These torsional backgrounds thus constitute a rather generic class of solution in the context of $\mathcal{N}=2$ compactifications.

A possible window in the far lesser known $\mathcal{N}=1$ heterotic flux compactifications is offered by orbifold constructions. Quotienting the previous flux compactification by a discrete group yields new heterotic solutions, which can break part of the symmetries of the initial vacuum~\cite{Becker:2008rc}. These are non-K\"{a}hler versions of the Borcea--Voisin construction of Calabi--Yau three-folds~\cite{Borcea:1996mxz,Voisin:1993ast}. Building this parent geometry from special classes of $K3$ surfaces\footnote{Such $K3$ surfaces are not generic in the moduli space.} which admit finite group actions with a non-trivial action on $(2,0)$ cohomology, it is possible to construct solutions which preserve only half of the initial supersymmetry, i.e. $\mathcal{N}=1$ in four dimensions. As $K3$ non-symplectic automorphisms generically admit a fixed locus, the resulting orbifold geometry inherits singular points or curves which await for resolution in order for the orbifold to define a consistent smooth supergravity solution (the singular geometry may still define a consistent string vacuum). However, while Calabi--Yau resolutions can be dealt with conveniently, notably through the machinery of toric geometry, carrying over these constructions to non-K\"{a}hler spaces is a challenging problem.

Fortunately, the worldsheet description of heterotic backgrounds can account for fluxes. A particularly fruitful approach is offered by some two-dimensional $(0,2)$ gauge theories, namely gauged linear sigma models (GLSMs), whose infrared fixed point is expected to provide non-linear sigma models on the desired target spaces~\cite{Witten:1993yc}. While initially introduced to describe Calabi--Yau compactifications, this construction has been extended 
in~\cite{Adams:2006kb} to  the $T^{2}$ bundles over $K3$ previously introduced. The key characteristic of this formulation is a worldsheet realization of the Green--Schwarz mechanism, which implements the $T^{2}$ fibration by an interplay between the chiral anomaly of the $K3$ base and a classical non-invariance of the torus fiber under gauge transformations. The resulting non-anomalous $(0,2)$ gauge theory is expected to flow in the infrared to a superconformal non-linear sigma model whose target space has the properties of the $\mathcal{N}=2$ flux solution. The theory also exhibits a non-anomalous $R$-symmetry, as well as a left-moving $U(1)$ flavor current, which is necessary to ensure the existence of an appropriate GSO projection.

This worldsheet construction provides a practical setting to extract several quantities about the compactifications. The massless spectra of the flux backgrounds can (in principle) be computed from a phase of the model which admits an asymmetric Landau--Ginzburg orbifold description \cite{Adams:2009zg}. Other aspects of the theory can also be explored, such as T-duality transformation properties \cite{Israel:2013hna} or computation of $\alpha'$-exact renormalization-invariant quantities as supersymmetric indices \cite{Israel:2015aea,Israel:2016xfu}, and the threshold 
corrections in the corresponding four-dimensional theory~\cite{Angelantonj:2016gkz}.

While constructed for flux backgrounds with $\mathcal{N}=2$ spacetime supersymmetry, this $(0,2)$ two-dimensional gauge theory model provides us with a very natural setting to study the $\mathcal{N}=1$ orbifolds 
of~\cite{Becker:2008rc}. Obtaining such orbifolds in a controlled worldsheet theory framework is essential. We stress again that a two-dimensional approach is unavoidable when it comes to heterotic flux compactifications, as the $\alpha'$ expansion of supergravity cannot be trusted, even more in the presence of singularities. Furthermore, with a $(0,2)$ description of the orbifolds, the previously mentioned worldsheet computations should carry over to the quotient field theories. Finally, recall that the $\mathcal{N}=1$ orbifolds of \cite{Becker:2008rc} generically admit a fixed locus. We will restrict ourselves to orbifolds which admit isolated fixed points only, keeping in mind their possible resolution. Indeed, while we expect the resolution of orbifolds with fixed curves to be rather involved, the situation is more 
familiar for orbifold with isolated fixed points only. In particular, there already exist well-developed techniques to handle the resolution of singularities in gauged linear sigma models \cite{GrootNibbelink:2010qut,Blaszczyk:2011hs}, so constructing a $(0,2)$ GLSM framework for these $\mathcal{N}=1$ orbifolds gives a first step in this direction.\footnote{As we will see later, the usual GLSMs constructions still need to be adapted to our particular setting, as the orbifolds considered here have a somehow non-standard action. In particular, they act by permutation on some vector and chiral multiplets.} Obtaining a resolved worldsheet model for $\mathcal{N}=1$ geometries would give a strong hint for the existence of $SU(3)$ structure backgrounds with fluxes, which would appear in the infrared limit of the worldsheet theory. This constitutes our main motivation for the following work.

This paper is organized as follows. In section~\ref{sec:flux_backgrounds}, we recall the conditions for  obtaining $\mathcal{N}=1$ backgrounds in heterotic supergravity, putting some emphasis on the torsional solutions and their extended supersymmetry. The worldsheet set up is reviewed in section~\ref{sec:GLSM}, with a description of $(0,2)$ gauged linear sigma-models and their torsional extension. With these in hand, section~\ref{sec:orbifolds} starts by a review of the main features of $\mathcal{N}=1$ orbifolds in the effective spacetime theory. Then, we introduce general properties of the corresponding orbifold in the worldsheet theory. As the orbifold of the GLSM yields a geometrical action in the low energy phase of the model, desired properties for this induced automorphism guide us to specific worldsheet models. In particular, a non-trivial orbifold action on vector multiplets is unavoidable. This results in strong constraints on the possible models. Vector multiplets have a direct interpretation in the geometrical phase, which guides us very naturally to selected geometries, namely hypersurfaces in products of projective spaces, or branched covers over those. After this general set up, we construct in 
sections~\ref{sec:Z2_orbifold} and~\ref{sec:Z3_orbifold} some explicit $\mathcal{N}=1$ orbifolds. These require the identification of distinguished configurations in the moduli space of $K3$ surfaces that satisfy all the constraints presented in the previous section. In section~\ref{sec:Z2_orbifold}, we introduce a smooth compactification, contructed from a discrete action of order two.  In section~\ref{sec:Z3_orbifold}, we present another orbifold, which results from a discrete $\mathbb{Z}_{3}$ action. The latter example admits singular points which will need to be resolved. Finally in section~\ref{sec:disc} we summarize the main result and suggest some avenues for future work.

\section{Heterotic compactifications and supersymmetry}
\label{sec:flux_backgrounds}
Let us start by reviewing supersymmetric four-dimensional flux compactifications of heterotic supergravity.

\subsection{Hull--Strominger equations}
The geometry of compactifications of the heterotic string to four dimensions is considerably constrained by supersymmetry. For a configuration preserving minimal supersymmetry, requiring the external space to be maximally symmetric leaves Minkowski as the only possibility. The internal six-dimensional space $X$ then has to come with a $SU(3)$ structure, which must solve the Hull--Strominger system \cite{Strominger:1986uh,Hull:1986kz}. For the hermitian $(1,1)$-form $J$ and the holomorphic $(3,0)$-form $\Omega$, this translates as the differential equations
\begin{equation}
\label{eq:flux_backgrounds:Strominger_system}
\text{d}\left(\text{e}^{-2\varphi}J\wedge J\right)=0,\qquad \text{d}\left(\text{e}^{-2\varphi}\Omega\right)=0.
\end{equation}
The bosonic fields are then obtained from this $SU(3)$ structure, with the internal metric being the hermitian metric associated to $J$, while the three-form flux $H$ follows from $H=\star\,\text{e}^{2\varphi}\text{d}(\text{e}^{-2\varphi}J)=\text{i}(\Bar{\partial}-\partial)J$. The dilaton is determined, up to a constant, by the exact Lee form of $J$ (and of $\Omega$).

Equations (\ref{eq:flux_backgrounds:Strominger_system}) are supplemented by equations for the gauge bundle $V$, which must be a holomorphic vector bundle whose corresponding curvature $F$ satisfies the Hermitian Yang--Mills equations with zero slope
\begin{equation}
F\wedge J^{2}=0,\qquad F\wedge\Omega=0,\qquad F\wedge\bar{\Omega}=0.
\end{equation}
Lastly, the gauge and tangent bundles are tied together by the Bianchi identity\footnote{The trace over gauge indices is normalized by fixing the index of embedding of the gauge group $G$ in $E_{8}\times E_{8}$ or $Spin(32)/\mathbb{Z}_{2}$.}
\begin{equation}
2\text{i}\partial\bar{\partial}J+\dfrac{\alpha'}{4}\left(\text{tr}F^{2}-\text{tr}R_{(+)}^{2}\right)=0,
\end{equation}
where $2\text{i}\partial\bar{\partial}J=\text{d}H$, and $R_{(+)}$ is the curvature of the connection with torsion $\Gamma_{(+)}=\Gamma+\tfrac{1}{2}H$. Provided this connection satisfies some $SU(3)$ instanton condition \cite{Ivanov:2009rh}, these supersymmetry conditions are enough to guarantee the full bosonic equations of motion, to first order in $\alpha'$.

\subsection{Heterotic flux backgrounds}
The non-K\"{a}hlerity of a heterotic background satisfying the Hull--Strominger system is encoded in the torsion $H=\text{i}(\bar{\partial}-\partial)J$. When this flux vanishes at leading order 
in $\alpha'$, the conditions for Calabi--Yau compactifications of~\cite{Candelas:1985en} are recovered. In this case, solutions are guaranteed by existence 
theorems~\cite{Yau:1978cfy,Donaldson:1985zz,Uhlenbeck:1986sl}, provided some topological conditions are satisfied: the K\"{a}hler space $X$ should have trivial canonical bundle, and the gauge bundle $V$ should be polystable. 

For non-K\"{a}hler backgrounds, the geometry is less controlled, but some classes of solutions are still known. The most trusted one was discovered in~\cite{Dasgupta:1999ss} by duality from type IIB orientifolds. 

\subsubsection*{Geometry of the internal space}
The geometry of the six-dimensional compactification space $X$ consists of a principal $T^{2}$ bundle over a $K3$ base $S$ with holomorphic two-form $\Omega_{S}$ and K\"{a}hler form $J_S$. Its $SU(3)$ structure, described in \cite{Goldstein:2002pg}, is constructed from the Calabi--Yau structure of the complex surface $S$ as
\begin{equation}
\label{eq:flux_backgrounds:SU(3)_structure}
J=\text{e}^{2\varphi}J_{S}+\dfrac{\text{i}\,a}{2\tau_{2}}\Theta\wedge\bar{\Theta},\qquad \Omega=\text{e}^{2\varphi}\sqrt{\dfrac{a}{\tau_{2}}}\,\Omega_{S}\wedge\Theta,
\end{equation}
where $a$ and $\tau=\tau_{1}+\text{i}\tau_{2}$ are the constant area and complex structure moduli, parametrizing the torus metric
\begin{equation}
\label{eq:flux_backgrounds:torus_metric}
\mathcal{G}_{IJ}=\dfrac{a}{\tau_{2}}\begin{pmatrix}
1&\tau_{1}\\
\tau_{1}&|\tau|^{2}\\
\end{pmatrix}.
\end{equation}
The $(1,0)$-form $\Theta$ appearing in (\ref{eq:flux_backgrounds:SU(3)_structure}) is built from two globally defined one-forms $\Theta^{I}$ as $\Theta=\Theta^{1}+\tau\Theta^{2}$. This one-form can be written locally as $\Theta=\text{d}\theta+A_{f}$, from the $T^{2}$ complex coordinate $\theta=\theta^{1}+\tau\theta^{2}$ and a complexified $U(1)$ connection one-form $A_{f}=A_{f}^{1}+\tau A_{f}^{2}$ on $K3$. Supersymmetry constrains the curvature $\omega_{f}=\frac{1}{2\pi\sqrt{\alpha'}}\text{d}\Theta=\omega_{f}^{1}+\tau \omega_{f}^{2}$ of this $T^{2}$ connection one-form\footnote{This normalization corresponds to a periodicity $\theta^{I}\sim \theta^{I}+2\pi\sqrt{\alpha'}$ for the torus coordinates.} as
\begin{equation}
\omega_{f}\wedge J_{S}=0,\qquad \omega_{f}\wedge \Omega_{S}=0,
\end{equation}
where $\omega_{f}^{I}\in H^{2}(S,\mathbb{Z})$. In particular, supersymmetry forbids a $(0,2)$ part for $\omega_{f}$, but turning on a $(2,0)$ part is still possible. However, such a piece is not present in the worldsheet construction considered in the following, and for our purposes $\omega_{f}$ will be considered to be a primitive $(1,1)$-form. We will come back later to the effect of this choice.

The flux corresponding to this geometry takes the form
\begin{equation}
\label{eq:flux_backgrounds:horizontal_vertival_decomposition}
H=\star_{S}\, \text{d}\text{e}^{2\varphi}+H_{I}\wedge\Theta^{I}
\end{equation}
with $H_{I}=-2\pi\sqrt{\alpha'}\,\mathcal{G}_{IJ}\,\omega_{f}^{J}$. This flux is subject to integrality restrictions \cite{Melnikov:2012cv,Melnikov:2014ywa}, which translate into quantization conditions $H_{I}\in H^{2}(S,2\pi\sqrt{\alpha'}\mathbb{Z})$\footnote{The difference with the quantization condition appearing in \cite{Melnikov:2014ywa} is due to a different normalization of torus coordinates.} for the forms $H_{I}$ appearing in the horizontal-vertical decomposition (\ref{eq:flux_backgrounds:horizontal_vertival_decomposition}). The resulting integrality constraint reads
\begin{equation}
\label{eq:flux_backgrounds:integrality_constraint}
\mathcal{G}_{IJ}\,\omega_{f}^{J}\in H^{2}(S,\mathbb{Z}).
\end{equation}
Consequently, the torus area is quantized. More generally, taking also into account a possible constant $B$-field along $T^{2}$, one finds that the complexified K\"{a}hler and complex structure moduli should belong to the 
same imaginary quadratic number field, see~\cite{Israel:2013hna} for details.

The gauge bundle is built out of a stable holomorphic bundle over $K3$ endowed with a Hermitian Yang--Mills connection, whose curvature satisfies
\begin{equation}
\label{eq:flux_backgrounds:susy_gauge_bundle}
F\wedge J_{S}=0,\qquad F\wedge\Omega_{S}=0,\qquad F\wedge\bar{\Omega}_{S}=0.
\end{equation}

Such a configuration respects all the conditions required by supersymmetry. The only missing equation is the Bianchi identity, which turns out to be a top form on $K3$. Indeed, the curvature $R_{(+)}$ is horizontal on $S$ when $\omega_{f}$ does not have a $(2,0)$ part \cite{Melnikov:2014ywa}. The Bianchi identity reads
\begin{equation}
\dfrac{1}{2}\Delta_{S}\,\text{e}^{2\varphi}\;J_{S}^{2}-\dfrac{4\pi^{2}\alpha'a}{\tau_{2}}\omega_{f}\wedge\bar{\omega}_{f}+\dfrac{\alpha'}{4}\text{tr}F^{2}-\dfrac{\alpha'}{4}\text{tr}R_{(+)}^{2}=0.
\end{equation}
In cohomology, it can be recast in the form 
\begin{equation}
\label{eq:flux_backgrounds:Bianchi_identity}
-\dfrac{a}{\tau_{2}}\omega_{f}\wedge\bar{\omega}_{f}-\dfrac{1}{2}p_{1}(V)+\dfrac{1}{2}p_{1}(TS)=0
\end{equation}
with $p_{1}(V)$ and $p_{1}(TS)$ the first Pontryagin classes of the gauge bundle and the $K3$ tangent bundle. Integration over the $K3$ base yields the consistency condition
\begin{equation}
\label{eq:flux_backgrounds:topological_condition}
-\dfrac{a}{\tau_{2}}\int\limits_{S}\omega_{f}\wedge\bar{\omega}_{f}+\dfrac{1}{16\pi^{2}}\int\limits_{S}\text{tr}F^{2}=24.
\end{equation}
Provided this topological condition is satisfied, it was shown in \cite{Fu:2006vj,Becker:2006et} that the Bianchi identity admits a smooth solution for the dilaton. The proof of~\cite{Fu:2006vj} is carried out with a different choice of connection for the tangent bundle, namely the Chern connection $\Gamma_{\text{C}}$. Indeed, for this connection the four-form $\text{tr}R_{\text{C}}^{2}$ is horizontal and $(2,2)$ ---~this is what allows a non-perturbative treatment of the Bianchi identity in the first place. However, existence of solutions is believed to still hold when using the connection with torsion~$\Gamma_{(+)}$, as the analysis of the partial differential equation does not rely too heavily on the details of the connection \cite{Melnikov:2014ywa}. Moreover, this non-perturbative analysis might even allow for a treatment of higher order $\alpha'$ corrections as perturbations of the non-linear solution.

\subsubsection*{Consequences of extended supersymmetry}
Discarding the $(2,0)$ piece for the complexified $T^{2}$ curvature $\omega_{f}$ has an important consequence. The $SU(3)$ structure of $X$ stays supersymmetric under any deformation of complex structure for the $K3$ base. This invariance under rotations of the hyper-K\"{a}hler structure of $K3$ reflects a greater amount of symmetry. Indeed, such a solution preserves $\mathcal{N}=2$ supersymmetry in spacetime, which is as much as $K3\times T^{2}$.

This difference can also be understood from an eight-dimensional perspective, by considering the heterotic theory compactified on $T^{2}$. There, it can be seen that a non-trivial fibration breaks some global worldsheet symmetries, whose nature is very different depending on the choice of $\omega_{f}$~\cite{Melnikov:2014ywa}. While turning on a $(1,1)$ part breaks a left-moving symmetry, tied to the gauge sector, a non-zero $(2,0)$ part would however break a right-moving symmetry, on the supersymmetric side of the heterotic string.

Extended supersymmetry leads to major simplifications on the supergravity side. The Bianchi identity becomes for $\mathcal{N}=2$ a single scalar equation on $K3$. Moreover, the torsional connection $R_{(+)}$ is purely horizontal. This leads to the hope that this equation can be corrected order by order in $\alpha'$, and that a perturbation of the solution of~\cite{Fu:2006vj} (which proves the existence of a solution to the Bianchi identity using the Chern connection for the tangent bundle) would solve this corrected equation up to any order in $\alpha'$. 

Another crucial difference between $\mathcal{N}=1$ and $\mathcal{N}=2$ appears when considering the worldsheet linear models of~\cite{Adams:2006kb}, described in the following section. Indeed, as a consequence of $(0,2)$ supersymmetry, the GLSM construction can accommodate $(1,1)$-forms only. This is the case in particular for the curvature of the torus bundle. Consequently, only the $\mathcal{N}=2$ solution admits a linear worldsheet description.

\section{Linear worldsheet models for flux compactifications}
\label{sec:GLSM}
A worldsheet construction for the torsional geometry described in the previous section can be formulated in the language of $(0,2)$ gauged linear sigma-models. In these models, introduced in~\cite{Witten:1993yc}, an explicit description of the internal space appears as some ``geometrical phase'', in the infrared limit. While the initial construction applied to Calabi--Yau compactification, the description of non-K\"{a}hler backgrounds with the topology of $T^{2}$ fibrations over a $K3$ base was considered in~\cite{Adams:2006kb}, with the generalization to arbitrary torus moduli described in~\cite{Israel:2013hna}. We summarize below the main features of the construction, for the abelian case which is relevant here.

\subsection{Torsional \texorpdfstring{$(0,2)$}{(0,2)} gauged linear sigma-model}
The six-dimensional heterotic background $X$ is described by a worldsheet Lagrangian of the form
\begin{equation}
\mathscr{L}=\mathscr{L}_{b}+\mathscr{L}_{g}+\mathscr{L}_{f},
\end{equation}
with $\mathscr{L}_{b}$ the contribution to the  Lagrangian from the base, $\mathscr{L}_{g}$ from the gauge bundle and $\mathscr{L}_{f}$ from the torus fiber. As for any chiral theory in two dimensions, the Lagrangian $\mathscr{L}$ could generically suffer from anomalies. Cancellation of these anomalies allows to link the base and the fiber through a Green--Schwarz type mechanism.

\subsubsection{Lagrangian for the \texorpdfstring{$K3$}{K3 } base}
The $K3$ base $S$ can be described as a $(0,2)$ gauged linear sigma-model, which is a two-dimensional gauge theory with $(0,2)$ supersymmetry. The field content of this model consists of a number $n_{v}$ of abelian vector multiplets $(\mathcal{A}^{\alpha}$, $\mathcal{V}^{\alpha}$), which couple to $n_{c}$ chiral multiplets $\Phi^{i}$, as well as to $n_{f}$ Fermi multiplets $\Gamma^{r}$. We denote by $q^{i}_{\alpha}$ and $q^{r}_{\alpha}$ the charges of the chiral and Fermi multiplets. Conventions for supersymmetry multiplets are spelled out in appendix~\ref{app:(0,2)susy}.

The Lagrangian for the base $S$ takes the form
\begin{equation}
\begin{split}
\mathscr{L}_{b}=&\int \text{d}^2\theta\left(-\frac{\text{i}}{2}\sum\limits_{i=1}^{n_{c}}\text{e}^{2q^{i}_{\alpha}\mathcal{A}^{\alpha}}\bar{\Phi}_{i} \mathscr{D}_{-}\Phi^{i}-\frac{1}{2}\sum\limits_{r=1}^{n_{f}}\text{e}^{2q^{r}_{\alpha}\mathcal{A}^{\alpha}}\bar{\Gamma}_{r}\Gamma^{r}-\frac{1}{8e^{2}}\delta_{\alpha\beta} \bar{\Upsilon}^{\alpha}\Upsilon^{\beta}\right)\\
&+\int\text{d}\theta\left(-\frac{\mu}{2}\sum\limits_{r=1}^{n_{f}}\Gamma_{r}J^{r} (\Phi)+\frac{1}{4}t_{\alpha}\Upsilon^{\alpha}\right) + \text{h.c.},
\end{split}
\end{equation}
with $\delta_{\alpha\beta}$ the Killing metric of the $U(1)^{n_{v}}$ algebra. The model is specified by the choice of polynomials $J^{r}(\Phi)$ appearing in the superpotential, and Fayet--Iliopoulos (FI)  parameters $t_{\alpha}=\text{i}r_{\alpha}+\frac{1}{2\pi}\theta_{\alpha}$. We assume that all the Fermi multiplets obey the standard chirality condition, i.e. $\bar{D}_{+}\Gamma = 0$. 

In the usual construction of~\cite{Witten:1993yc}, the GLSM is expected to admit an infrared limit where the gauge fields become non-dynamical\footnote{Formally, this limit corresponds to $e\to\infty$.} and can be integrated out. If the imaginary part of the FI  parameters is large, one gets a ``geometrical phase'' such that vacua of the model form a complex manifold: the scalars $\phi^{i}$ correspond to coordinates on a (weighted) projective space, and the vacuum manifold is carved into this ambient projective space by the polynomial equations $J^{r}(\phi)=0$. The massless right-handed fermions appearing in the chiral multiplets become sections of the tangent bundle, and the vectors correspond to $(1,1)$-forms in the target space, that we denote $\eta^{\alpha}\in H^{2}(S,\mathbb{Z})$. The sizes of the associated divisors are encoded in the imaginary part of the Fayet--Iliopoulos parameters. 

Several characteristic classes of the base $S$ can be read directly from the GLSM. In particular, the Chern characters read~\cite{Adams:2006kb}
\begin{subequations}
\begin{align}
&ch_{1}(TS)=\left(\sum\limits_{i=1}^{n_{c}}q^{i}_{\alpha}+\sum\limits_{r=1}^{n_{f}}q^{r}_{\alpha}\right)\eta^{\alpha},\\
&ch_{2}(TS)=\dfrac{1}{2}\left(\sum\limits_{i=1}^{n_{c}}q^{i}_{\alpha}q^{i}_{\beta}-\sum\limits_{r=1}^{n_{f}}q^{r}_{\alpha}q^{r}_{\beta}\right)\eta^{\alpha}\wedge\eta^{\beta}.
\end{align}
\end{subequations}
The Calabi--Yau condition $c_{1}(TS)=0$ for the vacuum manifold translates in the GLSM as the condition
\begin{equation}
\label{eq:GLSM:CalabiYau_condition}
\sum\limits_{i=1}^{n_{c}}q^{i}_{\alpha}+\sum\limits_{r=1}^{n_{f}}q^{r}_{\alpha}=0
\end{equation}
for the charges of the multiplets. One needs also to ensure that, for every Abelian worldsheet gauge field, the charges of chiral multiplets (including those appearing in the gauge bundle Lagrangian) sum to zero, in order that the FI couplings are invariant under RG-flow.\footnote{If these conditions are not satisfied for an otherwise consistent models, one can add ``spectator'' pairs of charged, massive chiral and fermi multiplets, without changing the IR geometry, see~\cite{Distler:1995mi}.}

The canonical example for the $K3$ base is built out of one vector multiplet $(\mathcal{A}$, $\mathcal{V})$, four chiral multiplets $\Phi_{1,\dots,4}$ of $U(1)$ charge $+1$ and one Fermi multiplet $\Gamma$ of charge $-4$, with the Fermat superpotential $J(\Phi)=\Phi_{1}^{4}+\dots+\Phi_{4}^{4}$. In the geometric phase, chiral fields get reinterpreted as $\mathbb{C}P^{3}$ homogeneous coordinates, and the superpotential carves out a quartic hypersurface $\mathbb{C}P^{3}[4]$ in this projective space. The vector multiplet gives rise to a $(1,1)$ form $\eta$ with self intersection $+4$. The K\"{a}hler form is $J_{S}=r\eta$, while the Chern classes are given by $c_{1}(TS)=0$ and $c_{2}(TS)=6\eta^{2}$.

Lastly, a crucial piece of the construction comes from the gauge anomaly of the Lagrangian. Indeed, as spinors of different chiralities live in different supersymmetry multiplets, their contributions to the variation of the path integral measure under a gauge transformation do not cancel out generically. The effective Lagrangian transforms as
\begin{equation}
\label{eq:GLSM:K3_anomaly}
\delta\mathscr{L}_{b,\,\text{eff}}=\dfrac{1}{8}\mathscr{A}_{b\,\alpha\beta}\int\text{d}^{2}\theta\;\Xi^{\alpha}\Upsilon^{\beta}+\text{h.c.}\qquad\text{with}\;\mathscr{A}_{b\,\alpha\beta}=\sum\limits_{i=1}^{n_{c}}q^{i}_{\alpha}q^{i}_{\beta}-\sum\limits_{r=1}^{n_{f}}q^{r}_{\alpha}q^{r}_{\beta}.
\end{equation}
While cancellation of any anomaly is necessary for the consistency of the full quantum theory, we do not require yet the vanishing of (\ref{eq:GLSM:K3_anomaly}), as the Lagrangian for the torus fiber will also give a non-zero (classical) contribution.

\subsubsection{Lagrangian for the gauge bundle}
The model can also accommodate a gauge bundle, which is described by supplementing a number $m_{f}$ of Fermi multiplets $\Psi^{s}$ of $U(1)$ charges $l^{s}_{\alpha}$, as well as $m_{c}$ chiral multiplets $P^{j}$ of charges $l^{j}_{\alpha}$.  The contribution to the Lagrangian describing the gauge bundle takes the form
\begin{equation}
\label{eq:GLSM:gauge_bundle_lagrangian}
\begin{split}
\mathscr{L}_{g}=&\int \text{d}^2\theta\left(-\frac{\text{i}}{2}\sum\limits_{j=1}^{m_{c}}\text{e}^{2 l^{j}_{\alpha}\mathcal{A}^{\alpha}}\bar{P}_{j} \mathscr{D}_{-}P^{j}-\frac{1}{2}\sum\limits_{s=1}^{m_{f}}\text{e}^{2l^{s}_{\alpha}\mathcal{A}^{\alpha}}\bar{\Psi}_{s}\Psi^{s}\right)\\
&-\frac{\mu}{2}\int\text{d}\theta\sum\limits_{j=1}^{m_{c}}P^{j}\left(\sum\limits_{s=1}^{m_{f}}\Psi^{s} Q_{js}(\Phi)\right) + \text{h.c.},
\end{split}
\end{equation}
where $Q_{js}(\Phi)$ are polynomials in the chiral multiplets of the base. In the geometric phase, the Fermi multiplets $\Psi^{s}$ correspond to sections of some gauge bundle built over the base $S$.

Similarly as before, Chern characters of the gauge bundle can be obtained straightforwardly from the charges of the multiplets as
\begin{subequations}
\begin{align}
&ch_{1}(V)=\left(\sum\limits_{j=1}^{m_{c}}l^{j}_{\alpha}+\sum\limits_{s=1}^{m_{f}}l^{r}_{\alpha}\right)\eta^{\alpha},\\
&ch_{2}(V)=\dfrac{1}{2}\left(-\sum\limits_{j=1}^{m_{c}}l^{j}_{\alpha}l^{j}_{\beta}+\sum\limits_{s=1}^{m_{f}}l^{s}_{\alpha}l^{s}_{\beta}\right)\eta^{\alpha}\wedge\eta^{\beta}.
\end{align}
\end{subequations}
(Notice the change of sign for $ch_{2}$ compared to the case of the base).

The Lagrangian for the gauge bundle is also generically anomalous, with the effective Lagrangian transforming as
\begin{equation}
\delta\mathscr{L}_{b,\,\text{eff}}=\dfrac{1}{8}\mathscr{A}_{g\,\alpha\beta}\int\text{d}^{2}\theta\;\Xi^{\alpha}\Upsilon^{\beta}+\text{h.c.}\qquad\text{with}\;\mathscr{A}_{g\,\alpha\beta}=\sum\limits_{j=1}^{m_{c}}l^{j}_{\alpha}l^{j}_{\beta}-\sum\limits_{s=1}^{m_{f}}l^{s}_{\alpha}l^{s}_{\beta}.
\end{equation}

The canonical example corresponding to the example of $\mathbb{C}P^{3}[4]$ described above consists of four Fermi superfields $\Psi^{s}$ of charge $+1$ and one chiral superfield $P$ of charge $-4$. The superpotential coupling is built out of polynomials $Q_{i}(\Phi)$ satisfying $\sum_{i=1}^{4}\Phi_{i}Q_{i}(\Phi)\propto J(\Phi)$. The case $Q_{i}=\partial_{i}J$ corresponds to the tangent bundle $TS$, while other choices give deformations of $TS$. In any case, the Chern classes of such a bundle are $c_{1}(V)=0$ and $c_{2}(V)=6\eta^{2}$.

\subsubsection{Lagrangian for the torus fiber}
The contribution to the Lagrangian describing the torus fiber is built out of two shift chiral multiplets $\Omega^{I=1,2}$ of integer charge $w^{I}_{\alpha}$. These are chiral superfields whose imaginary part shift-symmetry is gauged (rather than the phase of the superfield), see appendix~\ref{app:(0,2)susy} for details. In target-space, their (gauged) imaginary parts correspond to the real coordinates on the torus fiber, while their real parts are decoupled free bosons (which are an artifact of this construction). The Lagrangian takes the form
\begin{equation}
\label{eq:GLSM:torus_lagrangian}
\mathscr{L}_{f}=\dfrac{1}{2}\int\text{d}^2\theta\;\mathcal{G}_{IJ}\left(\text{Re}\,\Omega^{I}+w^{I}_{\alpha}\mathcal{A}^{\alpha}\right)\left(\partial_{-}\text{Im}\,\Omega^{J}+w^{J}_{\beta}\mathcal{V}^{\beta}\right)
-\dfrac{\text{i}}{4}h_{I\alpha}\int\text{d}\theta\;\Upsilon^{\alpha}\Omega^{I}+\text{h.c.},
\end{equation}
with the constant torus metric $\mathcal{G}_{IJ}$ defined by (\ref{eq:flux_backgrounds:torus_metric}). The superpotential couplings of the shift multiplets with the gauge super-curvatures $\Upsilon^{\alpha}$ are characterized by $h_{I\alpha} \in \mathbb{Z}$. These coupling constants are quantized in order for the path integral to be invariant under shifts $\Omega^{I} \mapsto \Omega^{I} + 2\text{i} \pi$ for any values of the instanton numbers $n^\alpha = -\tfrac{1}{2\pi} \int F^\alpha$. 

Only the imaginary part of the shift multiplets describe the relevant degrees of freedom of the torus, which can be seen from the invariance of the Lagrangian under integer shifts of $\text{Im}\,\Omega^{I}$. As supersymmetry requires the complexification of these coordinates into complex chiral multiplets, the non-compact part of the fiber described by $\text{Re}\,\Omega^{I}$ should decouple. This requirement translates into conditions on the couplings $h_{I\alpha}$ given by~\cite{Israel:2013hna}
\begin{equation}
h_{I\alpha}+\mathcal{G}_{IJ}w^{J}_{\alpha}=0.
\end{equation}
With this choice of axial couplings, the fiber Lagrangian describes a $\mathbb{C}\times T^{2}$ target space, where the non-compact direction is non-interacting and can be factorized out. In addition, quantization conditions on $h_{I\alpha}$ give back the integrality constraints (\ref{eq:flux_backgrounds:integrality_constraint}).

The Fayet--Ilioupoulos type gauge couplings of the shift multiplets with the gauge curvatures are classically anomalous. The non-invariance of the Lagrangian under a gauge transformation reads
\begin{equation}
\delta\mathscr{L}_{f}=-\dfrac{1}{4}\mathcal{G}_{IJ}w^{I}_{\alpha}w^{J}_{\beta}\int\text{d}\theta\;\Xi^{\alpha}\Upsilon^{\beta}+\text{h.c.}.
\end{equation}
In particular, the torus fiber can only add a negative contribution to non-mixed anomalies.

\subsection{Green–Schwarz mechanism}
In the previous construction, the base and the torus fiber live in different multiplets, and there is no interaction terms coupling them. Indeed, they only interplay with each other through their couplings to gauge fields. For the full theory to be consistent, gauge symmetry should be preserved in the effective theory. Under a gauge transformation, the Lagrangian varies as
\begin{equation}
\delta\mathscr{L}_{\text{eff}}=\dfrac{1}{8}\mathscr{A}_{\alpha\beta}\int\text{d}\theta\;\Xi^{\alpha}\Upsilon^{\beta}+\text{h.c.},
\end{equation}
with the shifted anomaly matrix
\begin{equation}
\label{eq:GLSM:anomaly_matrix}
\mathscr{A}_{\alpha\beta}=\mathscr{A}_{b\,\alpha\beta}+\mathscr{A}_{g\,\alpha\beta}-2\mathcal{G}_{IJ}w^{I}_{\alpha}w^{J}_{\beta}.
\end{equation}
The origin of this anomaly is very different for the base and the fiber. On one side, it is a one-loop effect due to spinors of different chiralities living in different $U(1)^{n_{v}}$ gauge representations. On the other side, it comes from a classical non-invariance of dynamical Fayet--Iliopoulos couplings in the action. Furthermore, this variation should not be corrected by higher order contributions in perturbation theory: the variation of the path integral measure is one-loop exact, and the superpotential is not renormalized beyond one-loop.

Vanishing of the anomaly matrix (\ref{eq:GLSM:anomaly_matrix}) has a direct interpretation in the geometric phase of the model. The equation can be rewritten at the level of forms, with $\omega^{I}_{f}=w^{I}_{\alpha}\eta^{\alpha}$, as
\begin{equation}
\dfrac{1}{2}\mathscr{A}_{\alpha\beta}\,\eta^{\alpha}\wedge\eta^{\beta}=
-\mathcal{G}_{IJ}\,\omega^{I}_{f}\wedge \omega^{J}_{f}-ch_{2}(V)+ch_{2}(TS).
\end{equation}
Vanishing of the anomaly reproduces exactly the Bianchi identity (\ref{eq:flux_backgrounds:Bianchi_identity}).

\section{Orbifolds with \texorpdfstring{$\mathcal{N}=1$}{N=1} spacetime supersymmetry}
\label{sec:orbifolds}
In the following, we describe the general construction of supersymmetric orbifolds of the worldsheet models described in section~\ref{sec:GLSM}.  We start by reviewing the supergravity construction introduced in~\cite{Becker:2008rc}, focusing on quotient manifolds that preserve four supercharges in spacetime.

\subsection{Discrete symmetry action and properties}
\label{subsec:N=1Orbifolds}
Starting from a heterotic flux background $X$ described in section~\ref{sec:flux_backgrounds}, which preserves $\mathcal{N}=2$ supersymmetry, new heterotic solutions have been constructed in~\cite{Becker:2008rc} by orbifolding $X$ by some discrete symmetry group $\mathbb{Z}_{p}$. Depending on the choice of orbifold, it is possible to preserve all supersymmetry, half of it, or to break it completely. We will consider orbifolds preserving $\mathcal{N}=1$ supersymmetry. We first recall the construction of~\cite{Becker:2008rc}.

Preservation of $\mathcal{N}=1$ supersymmetry under the orbifold restricts the discrete action on the $SU(3)$ forms $J$, $\Omega$ to be trivial. However, it is still possible to have a non-trivial action on the $K3$ base, as long as it is compensated by the orbifold action on the $T^{2}$ fiber. 

In order to preserve only half supersymmetry, the orbifold should break the hyper-K\"{a}hler structure of the base. Denoting by $\sigma$ a generator of $\mathbb{Z}_{p}$, the orbifold should act on the $K3$ surface $S$ as a non-symplectic automorphism
\begin{equation}
\sigma\ :\  J_{S}\to J_{S},\qquad\Omega_{S}\to\xi\,\Omega_{S},
\end{equation}
with $\xi=\text{e}^{2\pi \text{i}/p}$. This phase shift is balanced out by a rotation of the torus, written in local coordinates $\theta=\theta^{1}+\tau\theta^{2}$ as
\begin{equation}
\label{eq:N=1_orbifolds:action_torus_coordinate}
\sigma\ :\ \theta\to \bar{\xi}\,\theta.
\end{equation}
This rotation should be compatible with the lattice structure of the torus. For a generic complex structure parameter $\tau$, only the $\mathbb{Z}_{2}$ case is consistent, while specific points in moduli space also allow the possibility of $\mathbb{Z}_{3}$, $\mathbb{Z}_{4}$ and $\mathbb{Z}_{6}$. 

In order to accommodate the $T^{2}$-fibered structure of $X$, the $T^{2}$ connection one-forms also have to rotate under the orbifold as 
\begin{equation}
\label{eq:N=1_orbifolds:action_T2_connection}
\sigma\ :\ A_{f}\to\bar{\xi}\,A_{f}.
\end{equation}
This allows the globally defined form $\Theta$ to transform with the same phase rotation under the orbifold, and the holomorphic $3$-form $\Omega$ is thus invariant.

For later convenience, we rewrite this transformation in terms of real quantities as
\begin{subequations}
\begin{align}
\sigma\ :\ &\theta^{I}\to R(\sigma)^{I}{}_{J}\,\theta^{J},\\
&A_{f}^{I}\to R(\sigma)^{I}{}_{J}\,A_{f}^{J},
\end{align}
\end{subequations}
where $R(\sigma)$ denotes the rotation matrix
\begin{equation}
\label{eq:N=1_orbifolds:rotation_matrix}
R(\sigma)=\begin{pmatrix}
\text{Re}\,\xi+\frac{\tau_{1}}{\tau_{2}}\text{Im}\,\xi&\frac{|\tau|^{2}}{\tau_{2}}\text{Im}\,\xi\\
-\frac{1}{\tau_{2}}\text{Im}\,\xi&\text{Re}\,\xi-\frac{\tau_{1}}{\tau_{2}}\text{Im}\,\xi\\
\end{pmatrix}
\end{equation}
which preserves the torus metric (\ref{eq:flux_backgrounds:torus_metric}), as it obeys $R(\sigma)^{T}\mathcal{G}\,R(\sigma)=\mathcal{G}$.

Lastly, the action of the orbifold in the gauge sector must also be chosen. For the quotient geometry to be well-defined, the Yang--Mills curvature $F$ should be invariant under $\sigma$. A possibility
is to build $F$ out of forms on $S$ that are invariant under the discrete action. 
Different orbifold actions on gauge fields are also allowed, in particular it is possible to supplement the geometrical action of $\sigma$ on the $K3$ base with an action on the algebra generators. It is then possible to have a non-trivial action on the primitive forms of $S$ used to construct the gauge curvature, as long as a there is compensating action on the Lie algebra. Note that invariance of gauge fields is understood, of course, up to a gauge transformation.

If all of these requirements are met, the full supersymmetry constraints stay invariant under the orbifold, and the quotient geometry preserves $\mathcal{N}=1$ supersymmetry. 

\subsubsection*{Non-symplectic automorphisms of \texorpdfstring{$K3$}{K3}}
The orbifold action on the base $S$ corresponds to a  non-symplectic automorphism of $K3$. As these discrete $K3$ symmetries are quite constrained, only a limited number of cases need to be considered. We recollect here a few facts about such automorphisms, leading to the relevant possibilities for our orbifold construction. 

A non-symplectic automorphism of order $p$ of a $K3$ surface $S$ is a diffeomorphism $\sigma\,:\,S\to S$ which acts on the holomorphic $(2,0)$-form $\Omega_{S}$ as 
\begin{equation}
\label{eq:N=1_orbifolds:non_symplectic_automorphism}
\sigma^{\star}\,\Omega_{S}=\xi\,\Omega_{S},
\end{equation}
where $\xi$ is a $p$-th root of unity generating $\mathbb{Z}_{p}$. The automorphism $\sigma$ acts on the lattice $H^{2}(S,\mathbb{Z})$ by the isometry $\sigma^{\star}$, and its invariant sublattice is denoted $S(\sigma)$. As the holomorphic $(2,0)$-form is not invariant under the automorphism, $S(\sigma)$ has to be a sublattice of the Picard lattice~\cite{Nikulin:1980nk}
\begin{equation}
\text{Pic}(S)=H^{1,1}(S)\cap H^{2}(S,\mathbb{Z}).
\end{equation}

The fixed locus of a $K3$ non-symplectic automorphism can be either empty, or the disjoint union of points and smooth curves. This fixed locus can be completely characterized. Remarkably, for $p$ prime, the topological structure of the
fixed locus determines the action of $\sigma^{\star}$ on the $K3$ lattice $H^{2}(S,\mathbb{Z})$ uniquely~\cite{Artebani:2011kk}. 

For our purposes, as the orbifold should also act by a rotation of the torus fiber, the only automorphisms to consider are those compatible with the complex structure of the torus. For a generic complex structure parameter $\tau$, only an order two automorphism is consistent, while $\tau=\text{i}$ allows for the order four, and $\tau=\zeta=\text{e}^{2\pi\text{i}/3}$ allows for the orders three and six.\footnote{Other cases could be considered by acting on a $\Gamma_{2,2+n}$ lattice in place of the $\Gamma_{2,2}$ lattice of $T^{2}$.}

As we will explain later, the $K3$ automorphisms we want to consider should not leave invariant curves. Then the fixed locus of the automorphism will in most cases consist of a number of isolated fixed points. The existence of these invariant points  can be traced down to the holomorphic Lefschetz fixed-point formula, as underlined in~\cite{Becker:2008rc}. At any such point $p_{i}\in S$, the action of the automorphism $\sigma$ can be linearized and diagonalized to a local action $\sigma_{i}$. If $\sigma$ only admits isolated fixed points, the holomorphic Lefschetz formula then relates the local action at these points to the action $\sigma^{\star}$ on cohomology as
\begin{equation}
\sum\limits_{q}(-1)^{q}\text{tr}\big\rvert_{H^{0,q}(S)}\sigma^{\star}=\sum\limits_{\left\lbrace p_{i}\right\rbrace}\dfrac{1}{\text{det}(\mathbb{1}-\sigma_{i})}.
\end{equation}
For $\sigma$ a non-symplectic automorphism acting as (\ref{eq:N=1_orbifolds:non_symplectic_automorphism}) on $H^{2,0}(S)$, the traces in the left-hand side evaluate to $1+\bar{\xi}$. The Lefschetz formula then dictates the possible numbers of isolated fixed points, which differ depending on the order of the automorphism.

\paragraph{Order two}
For a $\mathbb{Z}_{2}$ action $\sigma\,:\,\Omega_{S}\to -\Omega_{S}$, the action of $\sigma$ at a fixed point can be locally diagonalized (up to a permutation of the coordinates) as  
\begin{equation}
\sigma_{i}=\begin{pmatrix}
-1&0\\
0&1\\
\end{pmatrix}
\end{equation}
and corresponds to a fixed curve: there are no isolated fixed points. This is compatible with the Lefschetz formula, as the trace of $\sigma^{*}$ in $H^{0}(S)$ and $H^{0,2}(S)$ gives respectively $+1$ and $-1$, so the total contribution is vanishing. Thus, the only possibility for of a non-symplectic involution without fixed curve is the Enriques involution, which acts freely on the surface $S$. 

\paragraph{Order three}
For a $\mathbb{Z}_{3}$ action $\sigma\,:\;\Omega_{S}\to\zeta\,\Omega_{S}$, the action of $\sigma$ can similarly be linearized and diagonalized at a fixed point to one of the following local actions
\begin{equation}
\sigma_{i}=\begin{pmatrix}
\zeta&0\\
0&1\\
\end{pmatrix},
\begin{pmatrix}
\zeta^{2}&0\\
0&\zeta^{2}\\
\end{pmatrix}.
\end{equation}
The first one corresponds to a fixed curve, while the second corresponds to an isolated fixed point. Such a point gives in the Lefschetz formula a contribution $\text{det}^{-1}(\mathbb{1}-\sigma_{i})=-\zeta/3$. As $\sigma$ acts by the identity on $H^{0}(S)$ and by $\zeta^{2}$ on $H^{0,2}(S)$, the contribution to the trace in cohomology is given by $1+\zeta^{2}=-\zeta$. Therefore, there is also a unique possibility for a non-symplectic automorphism without fixed curves~\cite{Artebani:2008ns}, which has exactly three fixed points.

\paragraph{Order four}
For an order four automorphism $\sigma\,:\,\Omega_{S}\to\text{i}\,\Omega_{S}$, one finds the following possible local actions
\begin{equation}
\sigma_{i}=\begin{pmatrix}
\text{i}&0\\
0&1\\
\end{pmatrix},
\begin{pmatrix}
-\text{i}&0\\
0&-1\\
\end{pmatrix}.
\end{equation}
Only the second one corresponds to an isolated fixed point, and has $\text{det}^{-1}(\mathbb{1}-\sigma_{i})=(1-\text{i})/4$. The cohomology contribution in the Lefschetz formula evaluates to $1-\text{i}$. Thus, any $\mathbb{Z}_{4}$ non-symplectic automorphism without fixed curve must have exactly four isolated fixed points.

\paragraph{Order six}
For an order six automorphism $\sigma\,:\,\Omega_{S}\to-\zeta^{2}\,\Omega_{S}$, the local action of the automorphism at any fixed point can be diagonalized to
\begin{equation}
\sigma_{i}=\begin{pmatrix}
-\zeta^{2} &0\\
0&1\\
\end{pmatrix},
\begin{pmatrix}
-\zeta&0\\
0&\zeta\\
\end{pmatrix},
\begin{pmatrix}
\zeta^{2}&0\\
0&-1\\
\end{pmatrix}.
\end{equation}
The last two cases correspond to isolated fixed points and have respectively $\text{det}^{-1}(\mathbb{1}-\sigma_{i})=(1-\zeta)/3$ and $\text{det}^{-1}(\mathbb{1}-\sigma_{i})=(1-\zeta)/6$. In the Lefschetz formula, the trace of $\sigma^{*}$ in Dolbeault cohomology gives a total contribution $1-\zeta$. Consequently, order six non-symplectic automorphisms with no fixed curves can have from three to six fixed points, depending on the local action of the automorphism at these points.
\paragraph{}
Thus, only few non-symplectic automorphisms of $K3$ can be considered depending on the discrete symmetry group: the Enriques involution for $\mathbb{Z}_{2}$, a non-symplectic automorphism with three fixed points for $\mathbb{Z}_{3}$, a non-symplectic automorphism with four fixed points for $\mathbb{Z}_{4}$, and non-symplectic automorphisms with three, four, five, or six fixed points for $\mathbb{Z}_{6}$. For all these automorphisms, the corresponding action on the lattice $H^{2}(S,\mathbb{Z})$ can be worked out, we refer to~\cite{Nikulin:1980nk,Artebani:2011kk} for details.

\subsection{Constraints from the worldsheet perspective}
In the torsional worldsheet model described in section~\ref{sec:GLSM}, the orbifold $\sigma$ becomes a discrete symmetry of order $p$ acting on $(0,2)$ multiplets. This symmetry should preserve the form of the Lagrangians $\mathscr{L}_{b}$, $\mathscr{L}_{g}$ and $\mathscr{L}_{f}$ independently. While in general, such $\mathbb{Z}_{p}$ symmetries would be quite unconstrained, here the discrete action must also reduce, in the geometric phase of the model, to the $\mathcal{N}=1$ orbifold discussed above. We will see that this restricts considerably possible classes of allowed models. 

\subsubsection{Orbifold action in the GLSM}
The orbifold action on the multiplets should both preserve the invariance of the three contributions to the Lagrangian independently and reproduce the required geometrical transformation in the geometric phase. For now we consider the general case of a $\mathbb{Z}_{p}$ orbifold $\sigma$, though cases of interest correspond to $p=2,3,4,6$.
\subsubsection*{Shift multiplets}
The shift chiral multiplets $\Omega^{I}$ correspond to torus coordinates, so they should transform accordingly as $\sigma\,:\,\Omega^{1}+\tau\Omega^{2}\to \bar{\xi}(\Omega^{1}+\tau\Omega^{2})$, or equivalently as
\begin{equation}
\label{eq:N=1_orbifolds:shift_multiplets}
\sigma\;:\;\Omega^{I}\to R(\sigma)^{I}{}_{J}\,\Omega^{J},
\end{equation}
with the rotation matrix $R(\sigma)$ defined in (\ref{eq:N=1_orbifolds:rotation_matrix}).

\subsubsection*{Vector multiplets}
Recall that for the orbifold action described in subsection~\ref{subsec:N=1Orbifolds}, the rotation (\ref{eq:N=1_orbifolds:action_torus_coordinate}) of the torus coordinates must be reproduced by the rotation of the $T^{2}$ connection one-forms, according to (\ref{eq:N=1_orbifolds:action_T2_connection}), in order for the quotient geometry to be defined consistently. Otherwise, the patching between different charts of $X$ would not be well-defined in the orbifold geometry. 

This consistency condition is mirrored in the GLSM. In eq.~(\ref{eq:GLSM:torus_lagrangian}), couplings between shift multiplets and vector fields prevent~(\ref{eq:N=1_orbifolds:shift_multiplets}) from being a discrete symmetry of the torus Lagrangian if vector fields stay invariant. The only option for $\mathscr{L}_{f}$ to be preserved by the orbifold is to allow for a non-trivial transformation of vector fields under $\sigma$. This transformation should be linear. We denote it by
\begin{subequations}
\label{eq:N=1_orbifolds:transformation_vectors}
\begin{align}
\sigma\;:\; &\mathcal{A}^{\alpha}\to \mathcal{S}(\sigma)^{\alpha}{}_{\beta}\,\mathcal{A}^{\beta},\\
&\mathcal{V}^{\alpha}\to \mathcal{S}(\sigma)^{\alpha}{}_{\beta}\,\mathcal{V}^{\beta},
\end{align}
\end{subequations}
with some constant matrix $\mathcal{S}(\sigma)$ satisfying $\mathcal{S}(\sigma)^{p}=I_{n_{v}}$. The condition for the invariance of the kinetic terms of the torus Lagrangian can be stated in terms of this matrix as 
\begin{equation}
\label{eq:N=1_orbifolds:invariance_vectors}
w^{I}_{\beta}\,\mathcal{S}(\sigma)^{\beta}{}_{\alpha}=R(\sigma)^{I}{}_{J}\,w^{J}_{\alpha}.
\end{equation}
This condition also guarantees the invariance of the axionic coupling appearing in eq.~(\ref{eq:GLSM:torus_lagrangian}).

Interpretation of $\mathcal{S}(\sigma)$ in the geometric phase is straightforward. Indeed, the $(1,1)$-forms $\eta^{\alpha}$ inherited from vector multiplets generate a sublattice of $\text{Pic}(S)$ of rank $n_{v}$. The rotation matrix $\mathcal{S}(\sigma)$ can be understood as the restriction of $\sigma^{\star}$ to this sublattice. Equation (\ref{eq:N=1_orbifolds:invariance_vectors}), obtained in the GLSM as a condition for invariance of the torus Lagrangian, amounts to the following transformation law of $\omega_{f}^{I}$
\begin{equation}
\sigma^{\star}\omega_{f}^{I}=R(\sigma)^{I}{}_{J}\,\omega_{f}^{J},
\end{equation}
or equivalently $\sigma^{\star}\omega_{f}=\bar{\xi}\omega_{f}$ for the complexified $T^{2}$ curvature.

Invariance of the gauge sector of the base Lagrangian under the orbifold gives additional conditions on the constant matrix $\mathcal{S}(\sigma)$ specifying the transformation of vector multiplets. In order for the gauge kinetic term to be unchanged, this matrix must preserve the Killing metric, i.e. it should be a $O(n_{v})$ matrix, such that 
$\mathcal{S}(\sigma)^{T}\delta\,\mathcal{S}(\sigma)=\delta$. Invariance of the Fayet--Ilioupoulos term is equivalent to the condition 
\begin{equation}
\label{eq:N=1_orbifolds:FayetIliopoulos_parameters}
t_{\beta}\,\mathcal{S}(\sigma)^{\beta}{}_{\alpha}=t_{\alpha}
\end{equation}
for the complexified K\"{a}hler parameters. Geometrically, this can be recast as the invariance of the K\"{a}hler form $\sigma^{\star}J_{S}=J_{S}$.

\subsubsection*{Chiral and Fermi multiplets}
For the matter sector, the orbifold action on chiral and Fermi multiplets must be specified. The discussion is very similar for both kinds of multiplets: the following comments for chiral multiplets can be straightforwardly translated to Fermi multiplets. 

Looking at the kinetic terms in the Lagrangians $\mathscr{L}_{b}$ and $\mathscr{L}_{g}$, there is an obvious possibility for the orbifold action on some chiral multiplet $\Phi$ of charge $q_{\alpha}$, which requires that the charges are invariant under the linear transformation of the vector multiplets, that is $q_{\beta}\,\mathcal{S}(\sigma)^{\beta}{}_{\alpha}=q_{\alpha}$. If this condition is satisfied, then any action of the form 
\begin{equation}
\sigma\;:\;\Phi\to \xi^{q}\Phi
\end{equation}
preserves the kinetic term. In this action, the orbifold acts by a phase rotation, with $q$ an integer and $\xi^{q}=\text{e}^{2\pi \text{i} q/p}$.

Such a choice turns out to be too restrictive, as invariance of the charges under $\mathcal{S}(\sigma)$ implies that some factor of the gauge group does not act on the chiral multiplets. There is a second possibility which allows chiral multiplets of non-invariant charges. For the orbifold action to leave the Lagrangian invariant, it is necessary that such chiral fields come in $p$-tuples $\Phi^{1},\Phi^{2},\dots,\Phi^{p}$, with the orbifold acting on the tuple by a permutation
\begin{equation}
\sigma\;:\; \Phi^{\ell}\to\Phi^{\ell+1}\qquad\text{for}\;\ell=1,\dots,p.
\end{equation}
(Here the notation is that $\Phi^{p+1}=\Phi^{1}$). The charges for such a $p$-tuple should be related by
\begin{equation}
\label{eq:N=1_orbifolds:charge_constraint}
q^{\ell+1}_{\alpha}=q^{\ell}_{\beta}\,\mathcal{S}(\sigma)^{\beta}{}_{\alpha}\qquad\text{for}\;\ell=1,\dots,p.
\end{equation}
Similar orbifold actions are possible for Fermi multiplets.

Choosing such an orbifold action on the chiral and Fermi multiplets, either with a phase or by a permutation, all the kinetic terms are invariant. The only transformation to worry about is the one of the superpotential. The orbifold $\sigma$ constrains greatly the choices of couplings and polynomials allowed in the superpotential. Some examples of models with invariant superpotentials will be discussed in later sections.

\subsubsection{Conditions for spacetime supersymmetry}
In the geometric phase that appears in the low energy limit of the GLSM (when the imaginary part of the FI parameters are large and positive), each one of the vector multiplets $(\mathcal{A}^{\alpha}$, $\mathcal{V}^{\alpha})$ gives a $(1,1)$-form $\eta^{\alpha}\in H^{2}(S,\mathbb{Z})$. The K\"{a}hler form of the base $S$ is constructed from these cohomology elements as
\begin{equation}
J_{S}=r_{\alpha}\,\eta^{\alpha},
\end{equation}
with $r_{\alpha}$ the imaginary parts of the FI parameters $t_{\alpha}$.

The two-form used to fiber the torus can be read from the Lagrangian $\mathscr{L}_{b}$ as
\begin{equation}
\omega_{f}=(w^{1}_{\alpha}+\tau w^{2}_{\alpha})\,\eta^{\alpha}.
\end{equation}
Recall that preservation of eight spacetime supercharges constrains this torus curvature $\omega_{f}$ to be a primitive form of $H^{1,1}(S)$. In the GLSM, $\omega_{f}$ is a $(1,1)$-form by construction, but the condition of primitivity is not generically satisfied. Therefore, this condition should be imposed on the models considered. In particular, there should be at least two vector multiplets. Otherwise, with only one vector multiplet, the resulting $J_{S}$ and $\omega_{f}$ would be proportional, and the corresponding background would not preserve spacetime supersymmetry.

Similarly, the gauge bundle is also constrained by supersymmetry to couple to the base with a primitive curvature $F\in H^{1,1}(S)$. In the GLSM, the gauge connection appears in the gaugings of Fermi multiplets in (\ref{eq:GLSM:gauge_bundle_lagrangian}). The choice of charges of Fermi multiplets is thus also constrained.

\subsubsection{Invariance of the anomaly matrix}
One crucial element in the worldsheet construction is the shifted anomaly matrix $\mathscr{A}_{\alpha\beta}$ associated with $U(1)^{n_{v}}$ gauge transformations of the GLSM, see eq.~(\ref{eq:GLSM:anomaly_matrix}). Gauge anomalies allow to link the $K3$ base, the gauge bundle and the torus fiber. Vanishing of $\mathscr{A}_{\alpha\beta}$ is a necessary condition for consistency of the quantum theory, just as the Bianchi identity is necessary for consistency of the supergravity solutions.

Being a symmetric matrix, the anomaly matrix has a priori $\frac{1}{2}n_{v}(n_{v}+1)$ components. Nevertheless, invariance of the Lagrangian under the orbifold implies that it obeys
\begin{equation}
\label{eq:N=1_orbifolds:anomaly_constraint}
\mathcal{S}(\sigma)^{\gamma}{}_{\alpha}\,\mathscr{A}_{\gamma\delta}\,\mathcal{S}(\sigma)^{\delta}{}_{\beta}=\mathscr{A}_{\alpha\beta}.
\end{equation}
This invariance of the anomaly matrix under $\mathcal{S}(\sigma)$ rotations ensures the vanishing of some of its components. One can choose a (complex) basis of the gauge algebra such that $S(\sigma)$ takes the diagonal form
\begin{equation}
\begin{pmatrix}
\mathbf{1}_{n_{0}}&&&\\
&\xi\mathbf{1}_{n_{1}}&&\\
&&\ddots&\\
&&&\xi^{p-1}\mathbf{1}_{n_{p-1}}\\
\end{pmatrix}
\end{equation}
where $n_{0},\dots,n_{p-1}$ are the dimensions of eigenspaces of $\mathcal{S}(\sigma)$, with $n_{0}+\dots+n_{p-1}=n_{v}$. Then, condition (\ref{eq:N=1_orbifolds:anomaly_constraint}) implies the vanishing of any mixed anomaly of two vector multiplets that do not sit in the same $S(\sigma)$ eigenspace. This guarantees the vanishing of some components of $\mathscr{A}_{\alpha\beta}$. The number of components left is $\frac{1}{2}n_{0}(n_{0}+1)+\dots+\frac{1}{2}n_{p-1}(n_{p-1}+1)$. These should be cancelled by choosing appropriate charges for the supersymmetry multiplets.

\subsection{Comments on compatible flux backgrounds}
In summary, the orbifold construction in the worldsheet theory is subject to several requirements:
\begin{enumerate}
\item the GLSM should, in some phase of its vacua, describe the topology of a smooth heterotic flux background $X$;
\item this background, before orbifolding, should preserve $\mathcal{N}=2$ supersymmetry in spacetime;
\item the action of orbifold on this geometric phase should preserve the $SU(3)$ structure of $X$, rotating the torus fiber along with a non-symplectic automorphism of the base;
\item the orbifold action should not admit any fixed curve in the vacuum manifold (though isolated fixed points are allowed).
\end{enumerate}
While the first three conditions only reflect the worldsheet analog of the orbifold construction of~\cite{Becker:2008rc}, the last one may seem more ad hoc. This condition stems from the fact that the singularities of the orbifold geometry need to be resolved in order to obtain a consistent supergravity background. For isolated fixed points, standard GLSM techniques for toric resolutions might apply to our $(0,2)$ description, in a similar fashion as the description of~\cite{GrootNibbelink:2010qut}.

Meeting all these requirements can, in practice, be difficult. The more constrained part turns out to be the Lagrangian for the base $S$. Indeed, by construction, $K3$ is realized in the GLSM as embedded in a product of projective spaces $\mathbb{C}P_{\alpha}$, with each vector multiplet $\Upsilon^{\alpha}$ giving rise to the associated K\"{a}hler form $\eta^{\alpha}$. The transformation law of vectors (\ref{eq:N=1_orbifolds:transformation_vectors}) thus dictates the kind of geometries one should be looking for: as vector multiplets are swapped by the orbifold, the same should go for the associated projective spaces. The base $S$ should therefore be constructed as a hypersurface in a product of projective spaces, and its non-symplectic automorphism should be inherited from a discrete action in the ambient space, which mixes the different $\mathbb{C}P_{\alpha}$. Branched covers of such hypersurfaces are also allowed by the GLSM construction, as we will see in the following sections.

Starting from such a model, some requirements of supersymmetry can be checked explicitly. As an example, consider a model for a $K3$ surface $S$ built out of $p$ vector multiplets $\Upsilon^{1},\Upsilon^{2},\dots,\Upsilon^{p}$, with the orbifold acting in the gauge sector by the permutation
\begin{equation}
\mathcal{S}(\sigma)=
\begin{pmatrix}
0&1&&&\\
&0&1&&\\
&&\ddots&\ddots&\\
&&&0&1\\
1&&&&0\\
\end{pmatrix}.
\end{equation}
For such a model, the only possible Fayet--Iliopoulos coupling for the base Lagrangian is of the form
\begin{equation}
\frac{1}{4}\int\text{d}\theta\,t\left(\Upsilon^{1}+\Upsilon^{2}+\dots+\Upsilon^{p}\right),
\end{equation}
from which the K\"{a}hler form of $S$ can be read as $J_{S}=r\left(\eta^{1}+\eta^{2}+\dots+\eta^{p}\right)$. In order to implement the $T^{2}$ fibration, a $(1,1)$-form with the correct transformation law under $\sigma$ can be obtained as $\omega_{f}=w\left(\eta^{1}+\xi\eta^{2}+\dots+\xi^{p-1}\eta^{p}\right)$, with $w\in\mathbb{R}$ corresponding to the charge of shift multiplets. The condition of primitivity for $\omega_{f}$ can then be checked directly, depending on the superpotential defining the hypersurface. 

Other examples can be built in a similar spirit. Nevertheless, it is not always guaranteed that such a geometry exists. Indeed, requiring a discrete $\mathbb{Z}_{p}$ symmetry restricts the possible choices for the superpotential. As there is a non-trivial action on the chiral multiplets, some monomials will not be allowed by the discrete
 symmetry. In some cases, the remaining allowed monomials will not be enough to define a smooth hypersurface: they may all vanish on a singular subset. 

Moreover, even in cases when such a smooth $\mathcal{N}=2$ background is found, with an orbifold mixing vector multiplets according to (\ref{eq:N=1_orbifolds:transformation_vectors}), it is still not sure that all constraints will be satisfied. Indeed, the transformation of $\Omega_{S}$ under the orbifold should also be checked, in order for the orbifold to break half supersymmetry. Finally, since we demand that the orbifold does not fix any curve of $S$, this reduces again the number of possible configurations.

\section{Smooth \texorpdfstring{$\mathcal{N}=1$}{N=1} compactifications from \texorpdfstring{$\mathbb{Z}_{2}$}{Z2} orbifolds}
\label{sec:Z2_orbifold}
In this section, we consider explicit worldsheet constructions for $\mathcal{N}=2$ heterotic flux backgrounds which admit a $\mathbb{Z}_{2}$ orbifold action that results in a freely-acting involution on the geometry, thus leading to smooth backgrounds preserving $\mathcal{N}=1$ spacetime supersymmetry.

\subsection{A model with two vector multiplets}
For the model we consider in the following, the $K3$ base is built out of six chiral multiplets that we denote $X_{1}$, $X_{2}$, $Y_{1}$, $Y_{2}$ and $W_{1}$, $W_{2}$, and two Fermi multiplets $\Gamma_{1}$, $\Gamma_{2}$. These fields couple to two vector multiplets $(\mathcal{A}^{1}$, $\mathcal{V}^{1})$ and $(\mathcal{A}^{2}$, $\mathcal{V}^{2})$, with charge assignment given in table~\ref{table:Z2orbifold-K3charges}. This specific choice of charges satisfies the Calabi--Yau condition (\ref{eq:GLSM:CalabiYau_condition}).
\begin{table}[h]
\begin{center}
\begin{tabular}{|c|cccccc|cc|}
\hline
&$X_{1}$&$X_{2}$&$Y_{1}$&$Y_{2}$&$W_{1}$&$W_{2}$&$\Gamma_{1}$&$\Gamma_{2}$\\
\hline
$U(1)_{1}$&$1$&$1$&$0$&$0$&$1$&$1$&$-2$&$-2$\\
$U(1)_{2}$&$0$&$0$&$1$&$1$&$1$&$1$&$-2$&$-2$\\
\hline
\end{tabular}
\end{center}
\caption{Charge assignment of the $\mathbb{Z}_{2}$ model}
\label{table:Z2orbifold-K3charges}
\end{table}

The superpotential that we consider is of the form
\begin{equation}
-\dfrac{\mu}{2}\int\mathrm{d}\theta\;\left(\Gamma_{1}\,J_{1}+\Gamma_{2}\,J_{2}\right),
\end{equation}
with $J_{1,2}(X,Y,W)$ two polynomials in the chiral multiplets of $U(1)_{1}\times U(1)_{2}$ charge $(2,2)$, to ensure classical gauge invariance. We also allow for two Fayet--Iliopoulos parameters $t_{1}$, $t_{2}$.

For the purposes of our construction, the resulting Lagrangian $\mathcal{L}_{b}$ should be invariant under a $\mathbb{Z}_{2}$ orbifold $\sigma$. This involution acts on the chiral multiplets as
\begin{equation}
\label{eq:Z2_orbifold:action_chiral}
\sigma\;:\;\left\lbrace X_{1},X_{2},Y_{1},Y_{2},W_{1},W_{2} \right\rbrace\to\left\lbrace Y_{1},Y_{2},X_{1},X_{2},-W_{1},-W_{2} \right\rbrace,
\end{equation}
while vector multiplets transform as
\begin{equation}
\label{eq:Z2_orbifold:action_vector}
\sigma\;:\;\left\lbrace (\mathcal{A}^{1},\mathcal{V}^{1}),(\mathcal{A}^{2},\mathcal{V}^{2})\right\rbrace\to\left\lbrace (\mathcal{A}^{2},\mathcal{V}^{2}),(\mathcal{A}^{1},\mathcal{V}^{1})\right\rbrace.
\end{equation}
Fermi multiplets $\Gamma_{1,2}$ stay invariant under the orbifold. As they appear in the superpotential, this  constrains possible choices for the polynomials $J_{1,2}$ which should also be invariant. An explicit example of such polynomials is
\begin{subequations}
\label{eq:Z2_orbifold:superpotential}
\begin{align}
J_{1}&=W^{2}_{1}+f_{1}(X,Y)\quad \text{with}\qquad f_{1}(X,Y)=X_{1}^{2}Y_{1}^{2}+X_{2}^{2}Y_{2}^{2}+\lambda_{1}\left(X_{1}^{2}Y_{2}^{2}+X_{2}^{2}Y_{1}^{2}\right),\\
J_{2}&=W^{2}_{2}+f_{2}(X,Y)\quad \text{with}\qquad f_{2}(X,Y)=X_{1}^{2}Y_{1}^{2}+X_{2}^{2}Y_{2}^{2}+\lambda_{2}\left(X_{1}^{2}Y_{2}^{2}+X_{2}^{2}Y_{1}^{2}\right),
\end{align}
\end{subequations}
where $\lambda_{1,2}$ are two complex constants. The last constraint imposed on the Lagrangian by requiring invariance under $\sigma$ is that the Fayet--Iliopoulos term should be of the form
\begin{equation}
\frac{1}{4}\int\text{d}\theta\,t\left(\Upsilon^{1}+\Upsilon^{2}\right),
\end{equation}
so only one Fayet--Iliopoulos parameter $t$ can be turned on in the orbifold theory.

In order to make contact with the constraints described in section~\ref{sec:orbifolds}, the action of $\sigma$ on the vector multiplets can be written in terms of the $O(2)$ matrix 
\begin{equation}
\label{eq:Z2_orbifold:action_vector_matrix}
\mathcal{S}(\sigma)=\begin{pmatrix}
0&1\\1&0\\
\end{pmatrix}.
\end{equation}
The charges of the two pairs of chiral multiplets $X_{1,2}$ and $Y_{1,2}$ satisfy the constraint (\ref{eq:N=1_orbifolds:charge_constraint}). The equality of the two Fayet--Iliopoulos parameters comes from the constraint (\ref{eq:N=1_orbifolds:FayetIliopoulos_parameters}).

\subsubsection*{Branched covering description}
The vacuum manifold in the ``geometrical phase'' corresponds to configurations of the fields for which the scalar potential vanishes. For the scalar components of the chiral multiplets, this amounts to six real equations: two of them come from D-terms, and take the form (in Wess--Zumino gauge)
\begin{subequations}
\begin{align}
|x_{1}|^{2}+|x_{2}|^{2}+|w_{1}|^{2}+|w_{2}|^{2}&=r,\\
|y_{1}|^{2}+|y_{2}|^{2}+|w_{1}|^{2}+|w_{2}|^{2}&=r.
\end{align}
\end{subequations}
The other four are F-terms and correspond to the vanishing of the polynomials $J_{1}$ and $J_{2}$:
\begin{subequations}
\label{eq:Z2_orbifold:F_terms}
\begin{align}
&w_{1}^{2}+f_{1}(x,y)=0,\\
&w_{2}^{2}+f_{2}(x,y)=0.
\end{align}
\end{subequations}
With the residual gauge symmetry acting by a phase on the scalars, the manifold described by these equations has real dimension four.

Notice that for $r>0$, $x_{1}$ and $x_{2}$ can never both vanish simultaneously. Moreover, they transform with the same charge under $U(1)_{1}$, which is complexified to $\mathbb{C}^{\star}$ by supersymmetry. Therefore, they can be understood as coordinates on a sphere $\mathbb{C}P^{1}$. Similarly, $y_{1}$ and $y_{2}$ correspond to coordinates on another sphere.

At a generic point of $\mathbb{C}P^{1}\times\mathbb{C}P^{1}$, two values of $w_{1}$ and two values of $w_{2}$ satisfy the F-term constraints (\ref{eq:Z2_orbifold:F_terms}), except at the branched locus defined by
\begin{subequations}
\begin{align}
&\Sigma_{1}\;:\;f_{1}(x,y)=0,\\
&\Sigma_{2}\;:\;f_{2}(x,y)=0.
\end{align}
\end{subequations} 
Both $\Sigma_{1}$ and $\Sigma_{2}$ correspond to a degree $(2,2)$ hypersurface in $\mathbb{C}P^{1}\times\mathbb{C}P^{1}$, so they are curves of genus $g(\Sigma_{1})=g(\Sigma_{2})=1$, which intersect at eight points. Both curves are smooth for the choice of $f_{1,2}$ given in (\ref{eq:Z2_orbifold:superpotential}) and generic choices of $\lambda_{1,2}$.\footnote{It is sufficient to choose $\lambda_{1}$ and $\lambda_{2}$ distinct and non-zero, with $\lambda_{1}^{2},\lambda_{2}^{2},\lambda_{1}\lambda_{2}\neq1$.} Thus, the vacuum manifold $S$ can be understood as a smooth $4$-fold cover of $\mathbb{C}P^{1}\times\mathbb{C}P^{1}$ branched over $\Sigma_{1}\cup\Sigma_{2}$. The Euler characteristic can be computed from
\begin{equation}
\chi(S)=4\,\chi(\mathbb{C}P^{1}\times\mathbb{C}P^{1})-2\,\chi(\Sigma_{1})-2\,\chi(\Sigma_{2})+\chi(\Sigma_{1}\cap\Sigma_{2}),
\end{equation}
which gives $\chi(S)=24$ as expected for a $K3$ surface. 

There are two distinguished classes $C^{1}$ and $C^{2}$ of $\text{Pic}(S)$, cut out by the two hyperplanes
\begin{subequations}
\begin{align}
&\left\lbrace p\right\rbrace\times \mathbb{C}P^{1},\\
&\mathbb{C}P^{1}\times\left\lbrace p\right\rbrace,
\end{align}
\end{subequations}
for some point $p$ of $\mathbb{C}P^{1}$. Their intersection matrix reads 
\begin{equation}
C^{\alpha}\cdot C^{\beta}=\begin{pmatrix}0&4\\4&0\\\end{pmatrix}.
\end{equation}
Denoting by $\eta^{1}$ and $\eta^{2}$ their Poincar\'{e} duals, the K\"{a}hler form of the $K3$ surface is $J_{S}=r(\eta^{1}+\eta^{2})$, while $\eta^{1}-\eta^{2}$ gives a primitive class with self intersection $-8$. In therms of these forms, the second Chern class of the $K3$ surface is $c_{2}(TS)=2(\eta^{1})^{2}+2(\eta^{2})^{2}+6\,\eta^{1}\eta^{2}$.

Note that these classes appear quite explicitly in the GLSM. Integrating out the vector multiplets in the infrared limit yields for example
\begin{equation}
\mathcal{A}^{1}-\mathcal{A}^{2}=-\dfrac{1}{2}\ln\left(\bar{X}_{1}X_{1}+\bar{X}_{2}X_{2}\right)+\dfrac{1}{2}\ln\left(\bar{Y}_{1}Y_{1}+\bar{Y}_{2}Y_{2}\right),
\end{equation}
with the Fubini--Study K\"{a}hler potential for both spheres.

\subsubsection*{Gauge bundle}
The geometry for the $K3$ base has to be supplemented by a gauge bundle $V$ over it. There are several possibilities to describe such a bundle. As the gauge bundle can only couple to primitive forms, the Fermi multiplets cannot be charged under $\Upsilon^{1}+\Upsilon^{2}$ (though couplings to $\Upsilon^{1}-\Upsilon^{2}$ are still allowed). Consequently, the anomaly associated to the K\"{a}hler form has to be cancelled by the chiral multiplets only. The choice of charges for these chiral multiplets then dictates possible couplings for the polynomials that appear in the superpotential.

A specific example is spelled out in table~\ref{table:Z2_orbifold:gauge_bundle}. This vector bundle over $S$ is described by four chiral multiplets and an even number $m_{f}=2N$ of Fermi multiplets. Only $2\,n_{\text{ins}}$ of these Fermi superfields are charged under gauge transformations, they will give contributions to the second Chern class $c_{2}(V)$. 

\begin{table}[h]
\begin{center}
\begin{tabular}{|c|cccc|cccc|}
\hline
&$P^{1}$&$P^{2}$&$P^{3}$&$P^{4}$&$\Psi^{1,\dots,n_{\text{ins}}}$&$\Psi^{n_{\text{ins}}+1,\dots,N}$&$\Psi^{N+1,\dots,N+n_{\text{ins}}}$&$\Psi^{N+n_{\text{ins}}+1,\dots,2N}$\\
\hline
$U(1)_{1}$&$-3$&$-1$&$0$&$0$&$1$&$0$&$-1$&$0$\\
$U(1)_{2}$&$0$&$0$&$-3$&$-1$&$-1$&$0$&$1$&$0$\\
\hline
\end{tabular}
\end{center}
\caption{Charge assignment for a gauge bundle $V$ over the $\mathbb{Z}_{2}$ model}
\label{table:Z2_orbifold:gauge_bundle}
\end{table}

The charge content does not fully define the fiber Lagrangian, as the superpotential must still be specified. The gauge charges given in table~\ref{table:Z2_orbifold:gauge_bundle} restrict possible couplings: for example, the Fermi multiplets $\Psi^{s=1,\dots,n_{\text{ins}}}$ can only couple with $P^{1}$ and $P^{2}$, through polynomials $Q_{1\,s}(X,Y)$ and $Q_{2\,s}(X,Y)$ of $U(1)_{1}\times U(1)_{2}$ charge $(2,1)$ and $(0,1)$. Similarly, $\Psi^{s=N+1,\dots,N+n_{\text{ins}}}$ can couple with $P^{3}$, $P^{4}$ through polynomials $Q_{3\,s}(X,Y)$ and $Q_{4\,s}(X,Y)$ of bidegree $(1,2)$ and $(1,0)$. 

For the corresponding vector bundle $V$, different choices for the superpotential correspond to different deformations. The topology of $V$ is still determined by the choice of charges of table~\ref{table:Z2_orbifold:gauge_bundle}, with Chern characters $ch_{1}(V)=-4(\eta^{1}+\eta^{2})$ and $ch_{2}(V)=-5((\eta^{1})^{2}+(\eta^{2})^{2})+2\,n_{\text{ins}}(\eta^{1}-\eta^{2})^{2}$. 

For this example, the orbifold action on $V$ takes the form
\begin{subequations}
\label{eq:Z2_orbifold:action_gauge}
\begin{align}
\sigma\ :\ &\left\lbrace P^{1},P^{2},P^{3},P^{4}\right\rbrace\to\left\lbrace P^{3},P^{4},P^{1},P^{2}\right\rbrace,\\
&\left\lbrace \Psi^{1},\dots,\Psi^{N},\Psi^{N+1},\dots,\Psi^{2N}\right\rbrace\to\left\lbrace \Psi^{N+1},\dots,\Psi^{2N},\Psi^{1},\dots,\Psi^{N}\right\rbrace.
\end{align}
\end{subequations}
Choosing polynomials $Q_{1\,s}(X,Y)=Q_{3\,s+N}(Y,X)$ and $Q_{2\,s}(X,Y)=Q_{4\,s+N}(Y,X)$, this $\mathbb{Z}_{2}$ transformation leaves the Lagrangian invariant.

The geometric interpretation of (\ref{eq:Z2_orbifold:action_gauge}) follows immediately: the gauge bundle is a sum of two factors $V=V_{1}\oplus V_{2}$, with $V_{1}$ built out of the $N$ first Fermi multiplets, and $V_{2}$ out of the $N$ last ones. These two factors are exchanged by the orbifold according to 
\begin{equation}
\label{eq:Z2_orbifold:action_gauge_factors}
\sigma\ :\ V_{1}\oplus V_{2}\to V_{2}\oplus V_{1}.
\end{equation}
This $\mathbb{Z}_{2}$ action can be understood as an exchange of the two $E_{8}$ factors in which $V$ is embedded, thereby reducing the rank of the gauge group, in the same spirit as in CHL constructions~\cite{Chaudhuri:1995fk}.

\subsubsection*{Fibering the torus}
Complementing the GLSM for $K3$ with a torus fiber is now straightforward. In order to preserve spacetime supersymmetry, the torus coordinates must be coupled to a primitive form. In our setting, this form is given freely by the vector multiplet $\Upsilon^{1}-\Upsilon^{2}$. Therefore, the only freedom in the torus Lagrangian is the choice of charges $w^{I}_{1}=-w^{I}_{2}$. These two charges carry the responsibility of cancelling the full anomaly of the base.

Fortunately, invariance under the $\mathbb{Z}_{2}$ orbifold constrains the allowed form of the anomaly matrix. The constraint (\ref{eq:N=1_orbifolds:anomaly_constraint}) along with the form (\ref{eq:Z2_orbifold:action_vector_matrix}) of the matrix $\mathcal{S}(\sigma)$ implementing the orbifold on the vector multiplets implies that the anomaly matrix must satisfy $\mathscr{A}_{11}=\mathscr{A}_{22}$, so only two components of the full anomaly have to be cancelled. One of these two can be compensated by the torus fiber, while for the other one, contributions from the base and the gauge bundle should balance out. For the choices of charges of the model, with the gauge bundle specified in table~\ref{table:Z2_orbifold:gauge_bundle}, this is indeed the case. Then for this choice, the full Lagrangian defines a consistent quantum theory on the condition that the charges $w^{I}_{1}=-w^{I}_{2}$ obey\footnote{Note that this condition is specific to the choice of bundle $V$.}
\begin{equation}
\frac{a}{\tau_{2}}\omega_{f}\wedge\bar{\omega}_{f}=(3-n_{\text{ins}})(\eta^{1}-\eta^{2})^{2}.
\end{equation}
Here, $n_{\text{ins}}=0,1,2$ allow for a non-trivial fibration, while for $n_{\text{ins}}=3$ one recovers $K3\times T^{2}$. For $n_{\text{ins}}>3$, it is not possible to construct a consistent theory.

We see in this example that the GLSM puts an upper limit $n_{\text{ins}}=3$ on the instanton number of the gauge bundle, very similar to the topological condition (\ref{eq:flux_backgrounds:topological_condition}) which arises in the classical theory. Indeed, Fermi multiplets in the gauge sector contribute to the anomaly with a similar sign than the shift multiplets associated to the torus fibration.

\subsubsection*{Enriques involution}
The action (\ref{eq:Z2_orbifold:action_chiral}) of $\sigma$ on the chiral multiplets can be understood geometrically in the $K3$ phase of the model. The two ambient spheres $\mathbb{C}P^{1}\times\mathbb{C}P^{1}$ are swapped by the involution. The same goes for the four covers defined by the values of $w_{1}$ and $w_{2}$. The transformation of the holomorphic form $\Omega_{S}$ under the involution can be worked out explicitly. This is most easily seen by embedding the surface $S$ in $\mathbb{C}P^{5}$ using the Segre map
\begin{equation}
\left[z_{1}:z_{2}:z_{3}:z_{4}:z_{5}:z_{6}\right]=\left[w_{1}:w_{2}:x_{1}y_{1}:x_{1}y_{2}:x_{2}y_{1}:x_{2}y_{2}\right].
\end{equation}
In these $\mathbb{C}P^{5}$ homogeneous coordinates, $S$ appears as a complete intersection $\mathbb{C}P^{5}[2,2,2]$ defined by the vanishing locus of three quadric polynomials
\begin{subequations}
\begin{align}
j_{1}&=z_{1}^{2}+z_{3}^{2}+z_{6}^{2}+\lambda_{1}\left(z_{4}^{2}+z_{5}^{2}\right),\\
j_{2}&=z_{2}^{2}+z_{3}^{2}+z_{6}^{2}+\lambda_{2}\left(z_{4}^{2}+z_{5}^{2}\right),\\
j_{3}&=z_{3}z_{6}-z_{4}z_{5}.
\end{align}
\end{subequations}
Choosing local coordinates of $\mathbb{C}P^{5}$, e.g. $u_{1}=z_{1}/z_{6}$,$\dots$,$u_{5}=z_{5}/z_{6}$, the holomorphic $(2,0)$-form has the local expression
\begin{equation}
\Omega_{S}=\dfrac{1}{u_{1}u_{2}}\text{d}u_{4}\wedge\text{d}u_{5}
\end{equation}
up to a constant factor. The involution acts on the $\mathbb{C}P^{5}$ coordinates as
\begin{equation}
\sigma\;:\;\left[z_{1}:z_{2}:z_{3}:z_{4}:z_{5}:z_{6}\right]\to\left[-z_{1}:-z_{2}:z_{3}:z_{5}:z_{4}:z_{6}\right].
\end{equation}
Translating this action in local coordinates, one can see that $\Omega_{S}$ transforms as $\sigma\ :\ \Omega_{S}\to -\Omega_{S}$, and the action of the orbifold on the base corresponds to a non-symplectic automorphism of $K3$.

It can be checked from the action (\ref{eq:Z2_orbifold:action_chiral}) that the involution $\sigma$ acts without fixed point: it is therefore the Enriques involution of the $K3$ surface $S$.

\subsection{A model with three vector multiplets}
The previous example has two vector multiplets, and the orbifold $\sigma$ acts on these multiplets by a permutation. This structure is not generic, and to avoid any misconception we give in the following a different example where the $\mathbb{Z}_{2}$ action on vector multiplets leaves one of them invariant. In the associated geometry, this invariant multiplet gives rise to an invariant primitive form, which can be coupled to the gauge sector.

\begin{table}[ht]
\begin{center}
\begin{tabular}{|c|ccccccc|cc|}
\hline
&$X_{1}$&$X_{2}$&$Y_{1}$&$Y_{2}$&$Z_{1}$&$Z_{2}$&$W$&$\Gamma_{1}$&$\Gamma_{2}$\\
\hline
$U(1)_{1}$&$1$&$1$&$0$&$0$&$0$&$0$&$1$&$-1$&$-2$\\
$U(1)_{2}$&$0$&$0$&$1$&$1$&$0$&$0$&$1$&$-1$&$-2$\\
$U(1)_{3}$&$0$&$0$&$0$&$0$&$1$&$1$&$0$&$-2$&$0$\\
\hline
\end{tabular}
\end{center}
\caption{Charge assignment of the second $\mathbb{Z}_{2}$ model}
\label{table:Z2orbifold-K3charges(2)}
\end{table}

The following example has $(0,2)$ field content spelled out in table~\ref{table:Z2orbifold-K3charges(2)}. This GLSM has three vector multiplets. The superpotential is chosen to be
\begin{subequations}
\begin{align}
&J_{1}=f(X,Y,Z)\ \text{with}\quad f(X,Y,Z)=(X_{1}Y_{1}+X_{2}Y_{2})(Z_{1}^{2}+Z_{2}^{2})+(X_{1}Y_{2}-X_{2}Y_{1})Z_{1}Z_{2},\\
&J_{2}=W^{2}+g(X,Y)\ \text{with} \quad g(X,Y)=(X_{1}^{2}+X_{2}^{2})(Y_{1}^{2}+Y_{2}^{2})+\lambda X_{1}X_{2}Y_{1}Y_{2},
\end{align}
\label{eq:Z2_orbifold:superpotential(2)}
\end{subequations}
with $\lambda$ a complex constant. Two Fayet--Iliopoulos parameters $t$ and $t'$ can also be turned on, with a contribution to the superpotential given by
\begin{equation}
\frac{1}{4}\int\text{d}\theta\, \big(t\left(\Upsilon^{1}+\Upsilon^{2}\right)+t'\,\Upsilon^{3}\big).
\end{equation}

The corresponding $K3$ surface can be understood as a double cover of $M$ over the curve $\Sigma$, where $M$ is the manifold defined by $f(X,Y,Z)=0$ with $\chi(M)=8$, while $\Sigma$ is the curve $f(X,Y,Z)=g(X,Y)=0$ with $g(\Sigma)=5$. Here $X_{1,2}$, $Y_{1,2}$ and $Z_{1,2}$ are understood as coordinates on $\mathbb{C}P^{1}\times\mathbb{C}P^{1}\times\mathbb{C}P^{1}$.

Fixing the coordinates on any one of the three $\mathbb{C}P^{1}$s defines a hyperplane. The associated classes $C^{1}$, $C^{2}$, $C^{3}$, have intersection numbers
\begin{equation}
C^{\alpha}\cdot C^{\beta}=
\begin{pmatrix}
0&4&2\\
4&0&2\\
2&2&0\\
\end{pmatrix}.
\end{equation}

The Enriques involution of the $K3$ surface descends from the orbifold 
\begin{subequations}
\begin{align}
\sigma\;:\;\left\lbrace X_{1},X_{2},Y_{1},Y_{2},Z_{1},Z_{2},W\right\rbrace&\to\left\lbrace Y_{1},Y_{2},X_{1},X_{2},-Z_{1},Z_{2},-W\right\rbrace,\\
\left\lbrace \Gamma_{1},\Gamma_{2}\right\rbrace&\to\left\lbrace \Gamma_{1},\Gamma_{2}\right\rbrace,\\
\left\lbrace (\mathcal{A}^{1},\mathcal{V}^{1}),(\mathcal{A}^{2},\mathcal{V}^{2}),(\mathcal{A}^{3},\mathcal{V}^{3})\right\rbrace&\to\left\lbrace (\mathcal{A}^{2},\mathcal{V}^{2}),(\mathcal{A}^{1},\mathcal{V}^{1}),(\mathcal{A}^{3},\mathcal{V}^{3})\right\rbrace.
\end{align}
\end{subequations}
It can be checked that this $\mathbb{Z}_{2}$ orbifold does not admit any fixed point.

The K\"{a}hler form of the $K3$ surface is $J_{S}=r\,(\eta^{1}+\eta^{2})+r'\,\eta^{3}$. The torus can couple to the primitive form $\eta^{1}-\eta^{2}$, that transforms with a sign under $\sigma$. In addition, the geometry also has a primitive invariant form that gauge Fermi superfields can couple to. This form is given by $r\,(\eta^{1}+\eta^{2})-(2r+r')\eta^{3}$.

\subsection{Heterotic/type II duality and orientifold}
In the lore of string duality, one of the most familiar examples is the relation between compactifications of the heterotic theory on $K3\times T^{2}$ and the type IIA theory on a $K3$-fibered Calabi--Yau manifold~\cite{Kachru:1995wm,Ferrara:1995yx}, related to the six-dimensional duality between type II on K3 and heterotic on $T^4$~\cite{Hull:1994ys}. For $T^{2}$ fibrations over $K3$, the presence of flux makes the duality less controlled. Potential duals were considered in~\cite{Melnikov:2012cv}, as some $K3$-fibered Calabi--Yau manifolds lacking a compatible elliptic fibration with section. 

Considering such a dual pair of a heterotic flux background $X$ and the corresponding $K3$-fibered Calabi--Yau $Y$, both preserving $\mathcal{N}=2$ supersymmetry, one can wonder how the heterotic orbifold would translate on the type II side under duality. The $\mathbb{Z}_{2}$ orbifold acts on the heterotic side without any fixed point. With an adiabatic argument similar to the one of~\cite{Vafa:1995gm}, one can expect this heterotic orbifold to be equivalent to some undetermined involution $\sigma_{Y}$ on the type II side. As this involution must preserve only $\mathcal{N}=1$ supersymmetry, it should be an orientifold, acting on the Calabi--Yau $Y$ by an orientation-reversing isometry. The resulting quotient geometries should yield a dual pair with $\mathcal{N}=1$ supersymmetry.

Such an $\mathcal{N}=1$ duality is actually discussed in~\cite{Vafa:1995gm} in the case of $K3\times T^{2}$. We expect that a similar reasoning still applies in the case of heterotic flux backgrounds, so we reproduce here the main arguments. The space $Y$ considered in~\cite{Vafa:1995gm} can be described as $K3$ fibered over $\mathbb{C}P^{1}$, and has a distinguished antiholomorphic involution, acting on the $\mathbb{C}P^{1}$ base coordinates $u$ with the fixed-point free involution $u\to -1/\bar{u}$ and preserving $K3$ fibers. Following the adiabatic argument, taking the $\mathbb{C}P^{1}$ base very large, a fiber-wise application of string duality leads to a heterotic compactification on $T^{4}$ fibered over $\mathbb{C}P^{1}$. This geometry can surely be understood as $T^{2}$ fibered over some $K3$ surface, with this $K3$ base described by an elliptic fibration. The antiholomorphic involution of $Y$ would then translate on the heterotic side to an inversion of the torus fiber coordinate, and an involution $\sigma_{S}$ of the base $S$. Describing $S$ as $T^{2}$ fibered over $\mathbb{C}P^{1}$, the adiabatic argument of above dictates how $\sigma_{S}$ should act: the $\mathbb{C}P^{1}$ base should transform with the same antiholomorphic involution as on the type IIA side, and $T^{2}$ fibers should be preserved. This freely acting symmetry of $S$ is the Enriques involution (with respect to some choice of complex structure). Thus, the construction found on the heterotic side reflects exactly the geometry described in previous sections.

While in the construction of~\cite{Vafa:1995gm} the orbifold action on the heterotic gauge bundle is less straightforward to derive, it is underlined in  that if the second Chern class is divided equally in the two $E_{8}$ factors, a possibility for the automorphism is an exchange of the two factors. This also reflects the orbifold action (\ref{eq:Z2_orbifold:action_gauge_factors}) found in the GLSM construction.

\paragraph{}
What can we learn about the type II dual of the heterotic $\mathcal{N}=1$ geometry described in this section ---~if such a dual exists? Based on the arguments of~\cite{Melnikov:2012cv}, the dual of a $\mathcal{N}=2$ heterotic flux background should be given by a Calabi--Yau space $Y$ admitting a $K3$ fibration, but no compatible elliptic fibration. In addition, this type II dual should admit a freely-acting antiholomorphic involution $\sigma_{Y}$ preserving the $K3$ fibration, so that $Y/\sigma_{Y}$ gives a $\mathcal{N}=1$ dual to $X/\sigma$ by the adiabatic argument.

If $Y$ is described by an embedding in some ambient manifold (e.g. a projective space or a toric variety), it seems reasonable to expect the involution to be inherited from an action on the ambient space. Whether $Y$ admits or not a freely-acting discrete automorphism group $\Gamma$ descending from symmetries of its ambient space can, in many cases, be checked explicitly. Some conditions allow ruling out manifolds, for example several topological indices of $Y$ (notably $\chi(Y)$) should be divisible by the order of the group $|\Gamma|$. For complete intersection Calabi--Yau manifolds (CICYs), configurations admitting free linear group actions have been investigated algorithmically in~\cite{Braun:2010vc}. Out of the $7890$ Calabi--Yau manifolds of~\cite{Candelas:1987kf}, only $166$ of them admit involutions inherited from a $\mathbb{Z}_{2}$ projective action. 
Such CICY configurations are especially simple and amenable to computations. Notably, most configurations admit a favorable description~\cite{Anderson:2017aux}, which is an embedding of the Calabi--Yau in a product of projective spaces, such that all divisors descend from the ambient space. Conditions for the existence of a $K3$/elliptic fibration (as well as their compatibility) can be framed as conditions on divisors, so for favorable CICY configurations they can be checked systematically. As an example, consider the Calabi--Yau space $Y$ constructed as a bidegree $(2,4)$ hypersurface in $\mathbb{C}P^{1}\times\mathbb{C}P^{3}$. It has $h^{1,1}(Y)=2$ and $h^{2,1}(Y)=86$, so it is a favorable configuration, with the two divisors $D$ and $D'$ descending from $\mathbb{C}P^{1}$ and $\mathbb{C}P^{3}$. The space $Y$ exhibits an obvious $K3$ fibration, with the corresponding divisor $D$  satisfying $D^{3}=D^{2}\cdot D'=0$ and $D\cdot c_{2}(Y)=24$. Consequently, $Y$ cannot admit a compatible elliptic fibration: this would require $h^{1,1}(Y)\geq3$.

In principle, the above critera limit the number of candidates for a $\mathcal{N}=1$ dual in a given class of geometries. However, one should be careful not to exclude too quickly manifolds that do not meet all requirements: sometimes Calabi--Yau spaces can admit discrete symmetries that do not descend from automorphisms of their ambient spaces. This phenomenon appears already for CICY spaces, where the same manifold can admit several configurations in different projective spaces, with the different descriptions related by ineffective splitting~\cite{Candelas:1987kf}. Sometimes different configurations share common projective discrete symmetries, but it can also happen that a symmetry is only manifest in one specific configuration. An instructive illustration of  this situation is presented in~\cite{Gray:2021kax}, where discrete symmetries of CICYs are studied using their favorable descriptions. In this setting, only $101$ manifolds admit a freely-acting involution, including some manifolds that were not in the original list of~\cite{Braun:2010vc}.

\section{Torsional compactifications with singularities from \texorpdfstring{$\mathbb{Z}_{3}$}{Z3} orbifolds}
\label{sec:Z3_orbifold}
For a $\mathbb{Z}_{3}$ orbifold, the generator $\sigma$, which acts on the holomorphic two-form as $\Omega_{S} \to \zeta \Omega_{S}$,  should act on the two-form $\omega_{f}$ used to fiber the torus as 
\begin{equation}
\label{eq:Z3_orbifold:T2_curvature_transformation}
\sigma\;:\;\omega_{f}\to\zeta^{2}\omega_{f}.
\end{equation}
The corresponding $K3$ base should thus have, in addition to its K\"{a}hler form, additional $(1,1)$-forms $\omega_{f}^{1}$ and $\omega_{f}^{2}$ whose $\mathbb{Z}_{3}$ transformation law reproduces (\ref{eq:Z3_orbifold:T2_curvature_transformation}). We introduce in this section a GLSM for a heterotic flux background $X$ admitting such a discrete symmetry. The orbifold is not freely-acting, but leaves nine points of $X$ fixed, based at three different points of $K3$. The quotient geometry is a singular supergravity background preserving $\mathcal{N}=1$ spacetime supersymmetry.

\subsection{A model with three vector multiplets}
A possible construction is the following: the $K3$ surface is constructed from seven chiral multiplets $X_{1}$, $X_{2}$, $Y_{1}$, $Y_{2}$, $Z_{1}$, $Z_{2}$, $W$, as well as two Fermi multiplets $\Gamma_{1}$, $\Gamma_{2}$. These superfields couple to three vector multiplets gauging a group $U(1)_{1}\times U(1)_{2}\times U(1)_{3}$, with charge assignment given in table~\ref{table:Z3orbifold-K3charges}. Again, this choice of charges satisfies the condition (\ref{eq:GLSM:CalabiYau_condition}) which ensures $c_{1}(TS)=0$.
\begin{table}[h]
\begin{center}
\begin{tabular}{|c|ccccccc|cc|}
\hline
&$X_{1}$&$X_{2}$&$Y_{1}$&$Y_{2}$&$Z_{1}$&$Z_{2}$&$W$&$\Gamma_{1}$&$\Gamma_{2}$\\
\hline
$U(1)_{1}$&$1$&$1$&$0$&$0$&$0$&$0$&$1$&$-1$&$-2$\\
$U(1)_{2}$&$0$&$0$&$1$&$1$&$0$&$0$&$1$&$-1$&$-2$\\
$U(1)_{3}$&$0$&$0$&$0$&$0$&$1$&$1$&$1$&$-1$&$-2$\\
\hline
\end{tabular}
\end{center}
\caption{Charge assignment of the $\mathbb{Z}_{3}$ model}
\label{table:Z3orbifold-K3charges}
\end{table}

The Lagrangian for the base is then fully specified by the polynomials appearing in the superpotential, which we will choose to be
\begin{subequations}
\label{eq:Z3_orbifold:superpotential}
\begin{align}
J_{1}=f(X,Y,Z)\qquad\quad&\text{with}\quad f(X,Y,Z)=X_{1}Y_{1}Z_{1}+X_{2}Y_{2}Z_{2},\\
J_{2}=W^{2}+g(X,Y,Z)\;&\text{with}\quad g(X,Y,Z)=X_{1}^{2}Y_{2}^{2}Z_{2}^{2}+\zeta^{2} X_{2}^{2}Y_{1}^{2}Z_{2}^{2}+\zeta X_{2}^{2}Y_{2}^{2}Z_{1}^{2}\nonumber\\
&\qquad\qquad\qquad\qquad +X_{2}^{2}Y_{1}^{2}Z_{1}^{2}+\zeta^{2} X_{1}^{2}Y_{2}^{2}Z_{1}^{2}+\zeta X_{1}^{2}Y_{1}^{2}Z_{2}^{2},
\end{align}
\end{subequations}
as well as the Fayet--Iliopoulos parameters, taken to be equal ($t_{1}=t_{2}=t_{3}$).

The coordinates $X_{1,2}$, $Y_{1,2}$ and $Z_{1,2}$ describe three $\mathbb{C}P^{1}$s. The degree $(1,1,1)$ polynomial $f(X,Y,Z)$ carves in this product a hypersurface $M$ of Euler characteristic $\chi(M)=6$. The geometry of the base $S$ can then be understood as a double cover of $M$, with the two branches corresponding to the two possible values of $W$. The branched locus is a curve $\Sigma$ of genus $g(\Sigma)=7$, corresponding to the vanishing of $g(X,Y,Z)$.  With the choice of superpotential (\ref{eq:Z3_orbifold:superpotential}), both $M$ and $\Sigma$ are smooth. Accordingly, the Euler characteristic is $\chi(S)=2\chi(M)-\chi(\Sigma)=24$.

Corresponding to the three vector multiplets, the geometry has three $(1,1)$-forms $\eta^{\alpha}$. These forms are the Poincar\'{e} duals of the classes $C^{\alpha}$ cut out by 
\begin{equation}
\left\lbrace p\right\rbrace\times\mathbb{C}P^{1}\times\mathbb{C}P^{1},\qquad \mathbb{C}P^{1}\times\left\lbrace p\right\rbrace\times \mathbb{C}P^{1},\qquad \mathbb{C}P^{1}\times\mathbb{C}P^{1}\times\left\lbrace p\right\rbrace.
\end{equation}
Their intersection matrix reads
\begin{equation}
C^{\alpha}\cdot C^{\beta}=
\begin{pmatrix}
0&2&2\\
2&0&2\\
2&2&0\\
\end{pmatrix}.
\end{equation}
The second Chern class of $S$ can be obtained directly from the anomaly matrix of the base Lagrangian and reads
\begin{equation}
c_{2}(TS)=(\eta^{1})^{2}+(\eta^{2})^{2}+(\eta^{3})^{2}+4(\eta^{1}\eta^{2}+\eta^{2}\eta^{3}+\eta^{3}\eta^{1}).
\end{equation}

\subsection{Orbifold and singular points}
The $\mathbb{Z}_{3}$ orbifold generator $\sigma$ acts on the $(0,2)$ multiplets as 
\begin{subequations}
\begin{align}
\sigma\;:\;\left\lbrace X_{1},X_{2},Y_{1},Y_{2},Z_{1},Z_{2},W\right\rbrace&\to\left\lbrace Y_{1},Y_{2},Z_{1},Z_{2},X_{1},X_{2},\zeta^{2}W\right\rbrace,\\
\left\lbrace \Gamma_{1},\Gamma_{2}\right\rbrace&\to\left\lbrace \Gamma_{1},\zeta^{2}\Gamma_{2}\right\rbrace,\\
\left\lbrace (\mathcal{A}^{1},\mathcal{V}^{1}),(\mathcal{A}^{2},\mathcal{V}^{2}),(\mathcal{A}^{3},\mathcal{V}^{3})\right\rbrace&\to\left\lbrace (\mathcal{A}^{2},\mathcal{V}^{2}),(\mathcal{A}^{3},\mathcal{V}^{3}),(\mathcal{A}^{1},\mathcal{V}^{1})\right\rbrace.
\end{align}
\end{subequations}
With the polynomials defining the hypersurface transforming as $\sigma\,:\, f\to f,\;g\to\zeta \,g$, the Lagrangian is invariant. In the geometric phase, the holomorphic two-form can be shown to transform as 
\begin{equation}
\sigma\;:\;\Omega_{S}\to\zeta\, \Omega_{S},
\end{equation}
and the orbifold $\sigma$ acts as a non-symplectic automorphism. It admits three fixed points corresponding to 
\begin{subequations}
\label{eq:Z3_orbifold:singular_points}
\begin{align}
&X_{1}=Y_{1}=Z_{1}=1,\\
&X_{2}=Y_{2}=Z_{2}=-1,-\zeta,-\zeta^{2},\\
&W=0
\end{align} 
\end{subequations}
(or any point in the same gauge equivalence class). 

This $K3$ Lagrangian can then couple to the torus Lagrangian, with complex structure parameter $\tau=\zeta$. In order to fiber the torus, the complexified torus curvature $\omega_{f}$ can be chosen proportional to the primitive $(1,1)$-form 
\begin{equation}
\eta^{1}+\zeta\eta^{2}+\zeta^{2}\eta^{3},
\end{equation}
such that it transforms as (\ref{eq:Z3_orbifold:T2_curvature_transformation}). This form can only compensate for some of the chiral anomalies of the GLSM, but the rest needs to be canceled by the gauge bundle. 
\ref{eq:N=1_orbifolds:charge_constraint}
Generically, any $U(1)^{3}$ GLSM could have six different anomalies (mixed and non-mixed). The invariance of the anomaly matrix of the orbifold (\ref{eq:N=1_orbifolds:anomaly_constraint}) forbids three of these to be non-zero. The coupling to the torus allows compensating for one of the three anomalies left, the other two must be cancelled by the gauge bundle. A possible choice is to take five charged chiral multiplets $P^{j}$ and $m_{f}$ Fermi multiplets $\Psi^{s}$, with charges spelled out in table~\ref{table:Z3Orbifold-GaugeBundle}. These couple to the base Lagrangian with the polynomials $Q_{js}(X,Y,Z)$ appearing in the superpotential, with the degree of the polynomials chosen to ensure gauge invariance, e.g. $Q_{1s}(X,Y,Z)$ should be of tridegree $(1,1,0)$. The resulting bundle $V$ has $ch_{1}(V)=-4(\eta^{1}+\eta^{2}+\eta^{3})$ and $ch_{2}(V)=-2((\eta^{1})^{2}+(\eta^{2})^{2}+(\eta^{3})^{2})-3(\eta^{1}\eta^{2}+\eta^{2}\eta^{3}+\eta^{3}\eta^{1})$. 
\begin{table}[h]
\begin{center}
\begin{tabular}{|c|ccccc|c|}
\hline
&$P_{1}$&$P_{2}$&$P_{3}$&$P_{4}$&$P_{5}$&$\Psi^{s}$\\
\hline
$U(1)_{1}$&$-1$&$0$&$-1$&$-1$&$-1$&$0$\\
$U(1)_{2}$&$-1$&$-1$&$0$&$-1$&$-1$&$0$\\
$U(1)_{3}$&$0$&$-1$&$-1$&$-1$&$-1$&$0$\\
\hline
\end{tabular}
\end{center}
\caption{Charge assignment for a gauge bundle $V$ over the $\mathbb{Z}_{3}$ model}
\label{table:Z3Orbifold-GaugeBundle}
\end{table}

The $\mathbb{Z}_{3}$ transformation for the gauge bundle must also be specified. As the fields $P_{1}$, $P_{2}$, $P_{3}$ carry different charges under the different $U(1)$ factors, they must be swapped by the orbifold. The two other multiplets $P_{4}$, $P_{5}$ as well as the Fermi multiplets stay invariant under $\sigma$. The polynomials $Q_{js}$ should then transform accordingly.

While for this example of gauge bundle, scalar charges do not sum to zero, non-renormalization of the FI parameters can still be obtained by adding a pair of spectator fields, as explained in~\cite{Distler:1995mi}. Supplementing the model with a chiral multiplet $S$ of charges $(1,1,1)$ and a Fermi multiplet $\Sigma$ of charges $(-1,-1,-1)$ does not change the anomaly nor the condition $c_{1}=0$. Moreover, adding a term $\int\text{d}\theta\,m_{s}S\Sigma+\text{c.c.}$ in the superpotential ensures that these spectators do not affect the geometric phase. These auxiliary multiplets are also compatible with the orbifold action: they have the same charge under the three $U(1)$ factors, so they can stay invariant under $\sigma$.

The resulting geometry has
\begin{equation}
\begin{split}
ch_{2}(TS)-ch_{2}(V)&=(\eta^{1})^{2}+(\eta^{2})^{2}+(\eta^{3})^{2}-(\eta^{1}\eta^{2}+\eta^{2}\eta^{3}+\eta^{3}\eta^{1})\\
&=(\eta^{1}+\zeta\eta^{2}+\zeta^{2}\eta^{3})\wedge(\eta^{1}+\zeta^{2}\eta^{2}+\zeta\eta^{3}).
\end{split}
\end{equation}
The full Lagrangian is then well-defined, provided the charges of the torus are chosen to cancel the corresponding anomaly. Quotienting by $\mathbb{Z}_{3}$ then gives a $\mathcal{N}=1$ geometry with singular points based at the three fixed points (\ref{eq:Z3_orbifold:singular_points}) of the $K3$ surface.

\section{Discussion}
\label{sec:disc}
In this work, we have extended the $(0,2)$ worldsheet description of $\mathcal{N}=2$ heterotic flux vacua, introduced in~\cite{Adams:2006kb}, to models preserving 
$\mathcal{N}=1$ supersymmetry in four dimensions, by considering $(0,2)$ gauged linear sigma-models admitting appropriate discrete symmetries. 

Supersymmetry constrains possible models, and in particular the classes of $K3$ surfaces that have to be considered. In the GLSM, the main takeaway is that the orbifold should have a non-trivial action on vector multiplets. These vector multiplets are usually associated to the symplectic forms of ambient projective spaces. In that respect, requiring the existence of a supersymmetry-preserving discrete symmetry in the GLSM guides us very naturally to selected geometrical settings: the torsional background must be constructed from a $K3$ base\footnote{As underlined above, hypersurfaces in weighted projective spaces or branched covers are also possible.} $S\subset\mathbb{C} P^{K_{1}}\times\dots\times\mathbb{C}P^{K_{n_{v}}}$, with vector multiplets spanning a rank $n_{v}$ sublattice of $\text{Pic}(S)$. In addition, this surface $S$ has to admit a discrete $\mathbb{Z}_{p}$ symmetry under which some $\mathbb{C} P^{K_{\alpha}}$ factors are permuted. 

Many distinct families of $K3$ surfaces can be constructed as complete intersections in products of projective spaces. Different configurations yield the same surface topologically, but they do not admit the same deformations. 
Indeed, while $K3$ admits a $20$-dimensional space of complex structure deformations, only a subspace of these moduli is accessible when describing it as a complete intersection. Consequently, not all classes of $K3$ surfaces admit projectively induced discrete symmetries. As an example, a smooth hypersurface of degree $(2,3)$ in $\mathbb{C} P^{1}\times\mathbb{C} P^{2}$ gives a $K3$ surface, and it can be shown that such a configuration does not admit a $\mathbb{Z}_{2}$ symmetry inherited from an involution of the ambient projective spaces~\cite{Candelas:1987kf}. In our construction, this configuration would not allow for a GLSM description of the Enriques involution. On the other hand, building $K3$ as a degree $(2,2,2)$  hypersurface in $\mathbb{C}P^{1}\times\mathbb{C}P^{1}\times\mathbb{C}P^{1}$ enables a linear $(0,2)$ description. 

For our purposes, a necessary feature of the construction is, more generally, a specific $K3$ class which admits a projectively induced $\mathbb{Z}_{p}$ symmetry, in order to lift the discrete action to the GLSM. Vector multiplets must transform under the orbifold so that they can couple to the torus shift multiplets, yielding a non-trivial fibration. Moreover, the resulting automorphism of the base $S$ must be non-symplectic, and should not leave invariant curves. While such constrains are obviously not satisfied by most members of  of the $K3$ moduli space, it is still possible to pick particular configurations that meet all requirements.

Orbifolds of order two can give smooth compactifications with minimal supersymmetry. However, such geometries only have $SU(2)\times\mathbb{Z}_{2}$ structure. For orbifolds of order three, four or six, the quotient geometry necessarily admits (at least) singular points, based at the points of $K3$ which are fixed by the non-symplectic automorphism. These orbifold singularities have to be resolved to obtain a consistent heterotic background. However, unlike the K\"{a}hler setting, there is no topological condition guaranteeing the existence of a solution. Thus, the resolved geometry must be obtained by smoothing out locally singular points, and patching these local solutions into a global compact geometry, while preserving the supersymmetry conditions~(\ref{eq:flux_backgrounds:Strominger_system}). No existence theorem for such resolutions has been 
proven so far.

The worldsheet construction should give a more workable framework in order to address the resolution. In $(0,2)$ GLSMs, resolution techniques~\cite{GrootNibbelink:2010qut,Blaszczyk:2011hs} typically rely on supplementing the model with an additional vector multiplet, along with a chiral multiplet gauged under this additional $U(1)$ symmetry. The resulting model has different phases, obtained by varying the Fayet--Iliopoulos parameter associated with the new vector. In some limit, the additional multiplets are set to zero and the singular orbifold reappears, with the discrete transformation given by a residual gauge transformation associated to the extra $U(1)$. In another limit, the model is fully resolved, and the new chiral multiplet becomes a coordinate on the divisor associated to the resolution. The GLSM then gives a model with the topology of the resolved manifold, which is expected to flow to a conformal field theory in the infrared.
Thus, obtaining a worldsheet model for the resolution would give strong evidence for the existence of a heterotic flux background with $SU(3)$ structure.

However, it should be noted that the orbifold action in our model is quite unusual: it does not act on chiral multiplets only by phase multiplication, but generically also requires permuting some of them for consistency. In addition, vector multiplets also transform with a constant rotation matrix. Adapting $(0,2)$ resolution methods thus requires some modifications: in particular, the orbifold can no longer be understood as a residual $U(1)$ gauge transformation. It would be interesting to see how to reframe the worldsheet resolutions in this context. A possible direction would be to understand them as non-abelian GLSMs: indeed, the orbifold action on $(0,2)$ multiplets can surely be understood as a residual non-abelian gauge transformation.

An indirect hint of the existence of a resolved geometry could also be found by studying  Landau--Ginzburg orbifold phases of these GLSMs, by a careful analysis of the twisted sectors. Indeed, while a freely acting orbifold would only project out sectors of the theory, for an orbifold with fixed points twisted operators also have to be considered. Resolution modes could then be identified as marginal operators in the twisted sectors of the superconformal field theories that the models flow to in the infrared. 

\section*{Acknowledgments}
We would like to thank Alessandra Sarti for useful correspondence, and Ruben Minasian for discussions.

\appendix 
\section{Conventions for \texorpdfstring{$(0,2)$}{(0,2)} superspace}
\label{app:(0,2)susy}
Covariance under $(0,2)$ supersymmetry is made manifest by working in $(0,2)$ superspace, built by supplementing the worldsheet coordinates $\sigma^{+},\sigma^{-}$ , with fermionic coordinates $\theta^{+}$, $\bar{\theta}^{+}$, where $\bar{\theta}^{+}=(\theta^{+})^{\dagger}$. The supersymmetry generators are then
\begin{equation}
Q_{+}=\dfrac{\partial}{\partial\theta^{+}} +\text{i}\bar{\theta}^{+}\partial_{+},\qquad \bar{Q}_{+}=-\dfrac{\partial}{\partial\bar{\theta}^{+}}-\text{i}\theta^{+}  \partial_{+},
\end{equation}
and satisfy the $(0,2)$ supersymmetry algebra
\begin{equation}
\left\lbrace\bar{Q}_{+},Q_{+}\right\rbrace=-2\text{i}\partial_{+}
\end{equation}
with $-\text{i}\partial_{+}$ the generator of the right-moving worldsheet translation. The supersymmetric derivatives are defined as
\begin{equation}
D_{+}=\dfrac{\partial}{\partial\theta^{+}} -\text{i}\bar{\theta}^{+}\partial_{+},\qquad \bar{D}_{+}=-\dfrac{\partial}{\partial\bar{\theta}^{+}}+\text{i}\theta^{+}  \partial_{+},
\end{equation}
and satisfy $\left\lbrace\bar{D}_{+},D_{+}\right\rbrace=2\text{i}\partial_{+}$.

\subsection{Supermultiplets}
Supersymmetry multiplets are constructed as $(0,2)$ superfields, which are complex functions $\Theta$ on superspace transforming under a supersymmetry variation as
\begin{equation}
\delta\Theta=\left(\epsilon^{+}Q_{+}+\bar{\epsilon}^{+}\bar{Q}_{+}\right)\Theta.
\end{equation}
A chiral multiplet $\Phi$ obeys the constraint
\begin{equation}
\bar{D}_{+}\Phi=0.
\end{equation}
Its component expansion is
\begin{equation}
\Phi=\phi+\sqrt{2}\theta^{+}\lambda_{+}-\text{i} \theta^{+}\bar{\theta}^{+}\partial_{+}\phi
\end{equation}
with $\phi$ a scalar and $\lambda_{+}$ a right-moving fermion.

A Fermi multiplet $\Gamma$ satisfies the same chirality constraint\footnote{This constraint can be generalized to $\bar{D}_{+}\Gamma=\sqrt{2}E(\Phi)$ with $E$ an holomorphic function of chiral fields.}
\begin{equation}
\bar{D}_{+}\Gamma=0,
\end{equation}
but its lowest component is a left-moving fermion $\gamma_{-}$. Its component expansion also displays an auxiliary field $G$ as
\begin{equation}
\Gamma =\gamma_{-}+\sqrt{2}\theta^{+}G-\text{i} \theta^{+}\bar{\theta}^{+}\partial_{+}\gamma_{-}.
\end{equation}

A vector multiplet is built out of two real superfields $\mathcal{A}$ and $\mathcal{V}$. Both are unconstrained but transform under super-gauge transformations as
\begin{equation}
\mathcal{A}\ \to\ \mathcal{A} +\dfrac{\text{i}}{2}(\bar{\Xi}-\Xi),\qquad\mathcal{V}\ \to\ \mathcal{V}-\dfrac{1}{2}\partial_{-}(\Xi +\bar{\Xi}),
\end{equation}
where the super-gauge parameter $\Xi$ is a chiral superfield. In order to make the field content of this multiplet more manifest, super-gauge transformations can be used to fix the Wess--Zumino gauge, where the vector multiplet has the component expansion
\begin{subequations}
\begin{align}
&\mathcal{A}=\theta^{+}\bar{\theta}^{+}A_{+},\\
&\mathcal{V}=A_{-}-2\text{i}\theta^{+}\bar{\mu}_{-}-2\text{i}\bar{\theta}^{+}\mu_{-}+2\theta^{+}\bar{\theta}^{+}D.
\end{align}
\end{subequations}
The residual gauge symmetry is then a usual $U(1)$ gauge transformation, generated by super-gauge parameters of the form $\Xi=\rho-\text{i} \theta^{+}\bar{\theta}^{+}\partial_{+}\rho$, with $\rho$ real. The gauge curvature appears in the Fermi multiplet
\begin{equation}
\Upsilon=\bar{D}_{+}(\partial_{-}\mathcal{A}+\text{i}\mathcal{V}),
\end{equation}
with the expansion
\begin{equation}
\Upsilon= -2 \left(\mu_{-}-\text{i}\theta^{+}(D-\text{i} F_{01})-\text{i}\theta^{+} \bar{\theta}^{+} \partial_{+} \mu_{-} \right)
\end{equation}
where $F_{01}=\frac{1}{2}F_{-+}$. 

\subsection{Gauge couplings}
Consider a $(0,2)$ model with a number $n_{v}$ of abelian vector multiplets $(\mathcal{A}^{\alpha}$, $\mathcal{V}^{\alpha})$. The gauged sector is coupled to matter by imposing the following gauge transformations for chiral superfields. A chiral multiplet $\Phi$ with charges $q_{\alpha}$ transforms under a super-gauge transformation as
\begin{equation}
\Phi\ \to\ \text{e}^{\text{i} q_{\alpha}\Xi^{\alpha}}\Phi.
\end{equation}
The derivative $\partial_{-}\Phi$ does not transform covariantly, and needs to be supplemented by the gauge-covariant derivative
\begin{equation}
\mathscr{D}_{-}\Phi=\partial_{-}\Phi+\text{i} q_{\alpha}(\mathcal{V}^{\alpha}-\text{i}\partial_{-}\mathcal{A}^{\alpha})\Phi
\end{equation} 
which has the right transformation law under super-gauge transformations, that is
\begin{equation}
\mathscr{D}_{-}\Phi\ \to\ \text{e}^{\text{i} q_{\alpha}\Xi^{\alpha}}\mathscr{D}_{-}\Phi.
\end{equation}
Note that the gauge curvature appears in 
\begin{equation}
\bar{D}_{+}\mathscr{D}_{-}\Phi=q_{\alpha}\Upsilon^{\alpha}\Phi.
\end{equation}

In the construction of our models, we also consider shift chiral multiplets which have different transformation laws. A shift multiplet $\Omega$ with charges $w_{\alpha}$ transforms under a super-gauge transformation as
\begin{equation}
\Omega \ \to\ \Omega+\text{i}w_{\alpha}\,\Xi^{\alpha}
\end{equation}

\subsection{Lagrangians with \texorpdfstring{$(0,2)$}{(0,2)} supersymmetry}
Throughout this paper, the action $S$ of a model is related to its Lagrangian $\mathscr{L}$ by
\begin{equation}
S=\frac{1}{2\pi}\int\text{d}^{2}\sigma\, \mathscr{L}.
\end{equation}
Manifestly $(0,2)$ supersymmetric Lagrangians are obtained by integrating superfields over their fermionic coordinates, with the Berezin measure normalized as
\begin{equation}
\int\text{d}\theta\;\theta^{+}=1,\qquad\int\text{d}^{2}\theta\;\theta^{+}\bar{\theta}^{+}=1.
\end{equation}
Here we consider the Lagrangian for a single vector multiplet $(\mathcal{A}$, $\mathcal{V})$, one chiral multiplet $\Phi$ of charge $q$ and one Fermi multiplet $\Gamma$ of charge $q'$, though generalization to multiple fields is straightforward.

The Lagrangian for the matter sector takes the form
\begin{equation}
\mathscr{L}_{m}=-\frac{\text{i}}{2}\int\text{d}^2\theta\;\text{e}^{2q\mathcal{A}}\,\bar{\Phi}\mathscr{D}_{-} \Phi\;-\dfrac{1}{2}\int\text{d}^2\theta\;\text{e}^{2q'\mathcal{A}}\,\bar{\Gamma}\Gamma\;-\dfrac{\mu}{2} \int\text{d}\theta\;\Gamma J(\Phi)-\dfrac{\mu}{2} \int\text{d}\bar{\theta}\;\bar{\Gamma} \bar{J}(\bar{\Phi})
\end{equation}
where the superpotential term is constructed from a holomorphic function $J$.\footnote{In our models this function will always be polynomial.} The Lagrangian of the gauge sector is
\begin{equation}
\mathscr{L}_{g}=-\dfrac{1}{8e^{2}}\int \text{d}^2 \theta \;\bar{\Upsilon}\Upsilon+\dfrac{t}{4}  \int\text{d}\theta\;\Upsilon+\dfrac{\bar{t}}{4}\int\text{d}\bar{\theta}\;\bar{\Upsilon}
\end{equation}
and includes a Fayet--Iliopoulos term, with $t=\text{i}r+ \frac{\theta}{2\pi}$.

The corresponding scalar potential (obtained by integrating out the auxiliary fields) is
\begin{equation}
U(\phi,\bar{\phi})=\dfrac{e^{2}}{8}\left(q|\phi|^2-r\right)^{2}+\dfrac{\mu}{2}|J(\phi)|^2.
\end{equation}

Lastly, the Lagrangian for a shift multiplet $\Omega$ of charge $w$ is of the form
\begin{equation}
\mathscr{L}_{s}=-\frac{\text{i}}{4} \int \text{d}^2 \theta \left( \Omega + \bar{\Omega}+2 w\mathcal{A} \right)
\left(\partial_{-} (\Omega-\bar{\Omega})+ 2\text{i} w \mathcal{V}\right)
\end{equation}

\bibliographystyle{JHEP}
\bibliography{Literature}
\end{document}